\documentclass[preprint,superscriptaddress,floatfix,preprintnumbers,nofootinbib,altaffilletter]{revtex4-1}
\usepackage{graphicx}
\usepackage{subfigure}
\usepackage{dcolumn}
\usepackage{bm}
\usepackage{amssymb}
\usepackage{color}
\usepackage[utf8]{inputenc}
\usepackage[OT1]{fontenc}
\usepackage{yfonts,amsmath,amsthm,amsfonts,amssymb,amscd, ulem}
\usepackage{enumerate}
\usepackage{fancyhdr}
\usepackage{mathrsfs}
\usepackage{cancel}
\usepackage{slashed}
\usepackage[flushleft]{threeparttable}
\usepackage[colorlinks=true, citecolor=purple, linkcolor=blue]{hyperref}
\usepackage{url}
\usepackage{makecell,booktabs}
\usepackage{braket}
\usepackage{relsize}
\usepackage{tikz-feynman}
\usepackage{multirow}

\begin{document}
\title{Heavy long-lived coannihilation partner from inelastic Dark Matter model and its signatures at the LHC}
\author{Jinhui Guo}
\email{guojh23@pku.edu.cn}
\affiliation{School of Physics and State Key Laboratory of Nuclear Physics and Technology, Peking University, Beijing 100871, China}

\author{Yuxuan He}
\email{heyx25@pku.edu.cn}
\affiliation{School of Physics and State Key Laboratory of Nuclear Physics and Technology, Peking University, Beijing 100871, China}

\author{Jia Liu}
\email{jialiu@pku.edu.cn}
\affiliation{School of Physics and State Key Laboratory of Nuclear Physics and Technology, Peking University, Beijing 100871, China}
\affiliation{Center for High Energy Physics, Peking University, Beijing 100871, China}

\author{Xiao-Ping Wang}
\email{hcwangxiaoping@buaa.edu.cn}
\affiliation{School of Physics, Beihang University, Beijing 100083, China}
\affiliation{Beijing Key Laboratory of Advanced Nuclear Materials and Physics, Beihang University, Beijing 100191, China}

\begin{abstract}
The coannihilation mechanism is a well-motivated alternative to the simple thermal freeze-out mechanism, where the dark matter relic density can be obtained through the coannihilation with a partner particle of similar mass with dark matter. When the partner particle is neutral, the inelastic nature of dark matter can help it to escape the direct detection limits. In this work, we focus on the coannihilation scenario in which the annihilation cross section is dominated by the partner-partner pair annihilation. We pay special interest on the parameter space where the coannihilation partner is long-lived, which leads to displaced signatures at the collider.
In such case, it opens the heavy mass parameter space for the coannihilation dark matter, comparing with those dominated by the partner-dark matter annihilation.
Specifically, we study an inelastic scalar dark matter model with a specific parameter space, which realizes the domination of partner-partner pair annihilation.
Then, we study two different realizations of the coannihilation partner decay and the existing constraints from the relic abundance, direct and indirect dark matter detection and the collider searches.
We focus on the channel that the long-lived coannihilation partner decays to dark matter plus leptons.
The high-luminosity LHC can reach good sensitivities for such heavy dark matter and coannihilation partner around 100--700 GeV.
\end{abstract}

\maketitle
\tableofcontents

\section{Introduction}
\label{sec:int}

The dark matter (DM) is a fundamental and unresolved problem of the particle physics, given the great triumph of the Standard Model (SM) in explaining the phenomenons observed in local laboratories and the astrophysical studies. The Weakly Interacting Massive Particle (WIMP) scenario is one of the most popular dark matter models, which can explain the dark matter relic abundance, $\Omega h^2 = 0.1198 \pm 0.0026$ \cite{Aghanim:2018eyx}, through its thermal freeze-out mechanism with a weak scale annihilation cross section. It hints new physics could be related with weak scale or higher. The scenario can be cross-checked using the large hadron collider (LHC), terrestrial direct searches of the DM particles and indirect searches for the DM annihilation products. Until now, dark matter escapes all the above searches and people start to think about alternatives.

The coannihilation mechanism is one of the possible alternatives \cite{Griest:1990kh}, where the dark matter
coupling to the SM particles can be quite small.
As a result, the dark matter pair annihilation cross section is small, which helps it to evade the strong constraints on the dark matter pair annihilation from the Cosmic Microwave Background (CMB)~\cite{Slatyer:2015jla, Planck:2018vyg} and the indirect searches~\cite{AMS:2014xys, AMS:2014bun, Fermi-LAT:2015att, Fermi-LAT:2016uux, DAMPE:2017fbg}. Due to the small coupling to SM particles, the direct searches constraints at deep underground experiments can also be safely evaded~\cite{LUX:2016ggv, CDEX:2019hzn, XENON:2018voc, PandaX:2021osp}.
In the coannihilation scenario, its relic abundance is obtained through the annihilation with a slightly heavier particle, denoted as the \textit{coannihilation partner}.
In general, it will decay back to the dark matter particle.
If the coupling and the mass splitting to dark matter are small enough, it can be a long-lived particle (LLP) at the detector scale~\cite{Alimena:2019zri}. Different from the DM, the coannihilation partner can have a sizable coupling to SM particles to obtain a large coannihilation cross section.
Therefore, it is possible for LHC to produce an abundance of the coannihilation partners. However, the detection might be difficult, for example, if the mass splitting is too small, the visible decay products of the coannihilation partner are too soft to detect.
If the coannihilation particle is charged under the SM gauge group, the partner can have significant interactions to the SM particles, such as in supersymmetric models~\cite{Jungman:1995df, Edsjo:1997bg, Ellis:1999mm} and in many simplified models~\cite{Baker:2015qna, Bertone:2016nfn, Buschmann:2016hkc, Bauer:2016gys, Buchmueller:2017uqu}.  The LHC can probe those charged coannihilation particles via disappearing tracks, which has been studied in Ref.~\cite{Khoze:2017ixx, Ambrogi:2018ujg}.

If the coannihilation partners are not charged under the SM gauge group, then they are \textit{neutral} coannihilation partners. The neutral partners could come from the same origin as the dark matter, for example the inelastic dark matter (iDM), coming from a degenerate mass spectrum and later splitting into two separate states~\cite{TuckerSmith:2001hy}. More specifically, the ultraviolet model starts with a complex scalar or Dirac fermion dark matter, which can be charged under a dark sector $U(1)_D$, and then splits into the dark matter state ${\rm DM_1}$ and the excited state ${\rm DM_2}$. The dark gauge boson $A'$ dominantly couples to ${\rm DM_1+DM_2}$, while the diagonal couplings to ${\rm DM_1 + DM_1}$ and ${\rm DM_2+ DM_2}$ are vanishing or suppressed by the small mass splitting~\cite{TuckerSmith:2001hy, Tucker-Smith:2004mxa}. These neutral coannihilation particles can be probed via the long-lived signatures, which has been done at Belle-II and LHC~\cite{Izaguirre:2015zva, Berlin:2018jbm, Duerr:2020muu, Kang:2021oes}. One can also look for them at future LHC, neutrino programs and fixed target experiments~\cite{Izaguirre:2017bqb, Berlin:2018jbm, Batell:2021ooj}.\footnote{It is also worth mentioning that the iDM with large mass splitting can be used to reopen the kinetic mixing dark photon parameter space for  $(\text{g-2})_\mu$ anomaly~\cite{Mohlabeng:2019vrz, Tsai:2019buq}.}

In this work, we study the LLP signatures from a scalar iDM model at the LHC. In our setup, we consider the dark matter and coannihilation particles coming from a complex scalar. The complex scalar can couple to SM Higgs directly through the scalar quartic coupling, which effect is less studied for coannihilation partner in the previous literature.
Previously, people usually focused on the Higgs portal dark matter for a singlet scalar dark matter or a complex scalar dark matter, with a scalar quartic coupling like $s^2 H^\dagger H $ or $S^* S H^\dagger H$~\cite{Silveira:1985rk, McDonald:1993ex, Burgess:2000yq, Davoudiasl:2004be, Barger:2007im, Barger:2008jx, Lerner:2009xg, Grzadkowski:2009mj}. Such dark matter model is heavily constrained by the direct detection experiments~\cite{Cline:2013gha, Feng:2014vea, Han:2015dua, Wu:2016mbe, Escudero:2016gzx, GAMBIT:2018eea, Athron:2018ipf}, especially the recent results from XENON1T~\cite{XENON:2018voc} and PandaX~\cite{PandaX-II:2017hlx, PandaX-II:2020oim, PandaX:2021osp}, leaving only the resonance region viable.
Different from Higgs portal dark matter model, a singlet coannihilation scalar will open the parameter space from DM direct detection~\cite{Ghorbani:2014gka, Casas:2017jjg, Coito:2021fgo, Maity:2019hre}, via significant coannihilation contribution.

In general, there are three kinds of coannihilation processes: ${\rm DM_1}+{\rm DM_1}$, ${\rm DM_1}+{\rm DM_2}$, and ${\rm DM_2}+{\rm DM_2}$.  Most of previous coannihilation studies~\cite{Izaguirre:2015zva, Izaguirre:2017bqb, Berlin:2018jbm, Duerr:2020muu, Kang:2021oes, Batell:2021ooj} focus on the coannihilation process ${\rm DM_1}+{\rm DM_2}$.\footnote{Ref.~\cite{DAgnolo:2018wcn} considered the process ${\rm DM_2}+{\rm DM_2}$ but only for very light DM.} In this case, a small coupling between DM and coannihilation partner is necessary to make the partner long-lived. Therefore, one has to lower the DM mass scale to compensate this small coupling for the relic abundance. As a result, the DM mass has to be lighter than 100 GeV.
However, our coannihilation partner couples to SM Higgs via the scalar quartic, we can have a large $({\rm DM_2}+{\rm DM_2})$ partner pair annihilation cross section ($\sigma_{22}$). Later, we will build an ultraviolet model for the specific quartic coupling from a broken symmetry.
In our setup, the coannihilation partner (${\rm DM_2}$) couples to the SM Higgs, while the dark matter (${\rm DM_1}$) does not couple to SM Higgs directly, which is different to the Higgs portal DM model.
In our model, the DM pair annihilation cross section $\sigma_{11}$ is vanishing. The DM-partner annihilation cross section $\sigma_{12}$ is sub-dominant in the contribution of relic abundance, which separates our study from the previous ones. Since the relic abundance is fulfilled by the coannihilation partner pair annihilation, we can focus on much larger dark matter mass region ($> 100$ GeV), where the decay products are much more energetic than light DM scenario.
As a result, in our work, the annihilation channels for relic abundance, production channel at collider and DM mass region are quite different from the previous studies. Next we study the existing constraints for this model from collider, direct and indirect searches. 
Later, we will study an ultraviolet model in a specific parameter space, which leads to a special quartic coupling.

We organize the paper as follows. In section~\ref{sec:idm}, we describe the scalar inelastic dark matter models and the possible decay channels for the long-lived coannihilation partner. In section~\ref{sec:excon}, we discuss the existing constraints from dark matter relic abundance, direct detection, indirect detection and collider searches. In section~\ref{sec:LLP}, we discuss the long-lived particle signatures of the coannihilation partner and its detection at the LHC. In section~\ref{sec:conclusions}, we conclude.

\section{The Models}
\label{sec:idm}

The coannihilation mechanism can contribute significantly to the DM relic abundance. For this purpose, the coannihilation partner number density should be comparable to the DM. As a result, its mass can not be too large comparing with DM. In our study, we consider a complex scalar iDM model, with the real scalar ground state $s_1$ and excited state $s_2$ as the coannihilation partner. The dimensionless mass splitting between $s_1$ and $s_2$ is defined as
\begin{align}
\Delta \equiv \frac{m_2 - m_1 }{m_1},
\end{align}
where $m_{1,2}$ are the mass for $s_{1,2}$. If assuming the density ratio between $s_1$ and $s_2$ follows the equilibrium value,  one can solve the Boltzmann equation and obtain an effective cross section~\cite{Griest:1990kh, Baker:2015qna,Edsjo:1997bg,Bell:2013wua}
\begin{align}
\label{eq:sigma}
\sigma_{\text{eff}} = \frac{g^2_{s_1}}{g^2_{\rm eff}}\left(\sigma_{1 1} + 2 \sigma_{1 2} \frac{g_{s_2}}{g_{s_1}} (1+\Delta)^{3/2} e^{-x\cdot \Delta} + \sigma_{2 2} \frac{g^2_{s_2}}{g^2_{s_1}} (1+\Delta)^{3} e^{-2 x\cdot \Delta} \right),
\end{align}
where $\sigma_{i j }=\sigma(s_i~s_j \rightarrow \mathrm{SM~SM})$ is the annihilation cross section to SM particles,  $g_{s_1} = g_{s_2} = 1$ are the degrees of freedom for real scalar $s_1$ and $s_2$, and $x = m_\mathrm{DM}/T$ where $T$ is the temperature of thermal bath. The effective degree of freedom $g_{\rm eff}$ is defined as
\begin{equation*}
    g_{\text{eff}} = g_{s_1} + g_{s_2} (1+ \Delta)^{3/2} e^{-x\cdot \Delta }.
\end{equation*}

When the cross section $\sigma_{11}$ is negligible, the dominant contributions to effective cross section $\sigma_{\text{eff}}$ come from the coannihilation. The previous studies focused on the case that $\sigma_{12}$ is the dominant contribution to the effective annihilation cross section. We consider an alternative case that the coannihilation DM model leads to the following annihilation cross section,
\begin{align}
	\sigma_{1 1} \approx 0 , \quad \sigma_{1 2} \ll \sigma_{2 2}.
	\label{eq:coannihilation-crosssection}
\end{align}
It can enable us to consider the heavy DM parameter space and more energetic decay objects from long-lived $s_2$. The concrete model satisfying this feature will be introduced in the following subsection.

\subsection{Inelastic scalar dark matter model}

We start with the Lagrangian for a massive complex scalar $\hat{S}$, which satisfies a global $U(1)$ symmetry,
\begin{align}
    \mathcal{L}_{U(1)} & =  \left(\partial_\mu \hat{S}\right)^*  \left(\partial^\mu \hat{S} \right) - m_S^2 \hat{S}^* \hat{S},
\end{align}
where $\hat{S} = \left(\hat{s}_1 + i \hat{s}_2 \right)/\sqrt{2}$ is a complex scalar and $\hat{s}_{1,2}$ are the real scalars. The notation with a hat, e.g. $\hat{S}$, is for flavor eigenstates, and we reserve the notation without a hat for mass eigenstates. Then we add a quadratic term $\hat{s}_i \hat{s}_j$ into the Lagrangian $\mathcal{L}_\text{\sout{U(1)}} $ to explicitly break the $U(1)$ symmetry:
\begin{align}
    \mathcal{L}_\text{\sout{U(1)}} & = - \delta \hat{m}^2_{ij} \hat{s}_i \hat{s}_j - \hat{\lambda}_{ij} \hat{s}_i \hat{s}_j \left(H^\dagger H -\frac{v^2}{2}\right),
\end{align}
where $H$ is the SM Higgs doublet and $v$ is the SM Higgs vacuum expectation value (vev). The mass matrix $\delta \hat{m}^2$ and scalar quartic coupling matrix $\hat{\lambda}$ are real symmetric matrices. We neglect other self-interacting quartic scalar terms which are irrelevant in this work.

To obtain the mass eigenstates, one can apply a $U(1)$ rotation $U$, parameterized with an angle $\theta$,
\begin{equation}
	U = \begin{pmatrix}
		\cos \theta & \sin \theta \\
		- \sin \theta &\cos \theta
	\end{pmatrix},
\end{equation}
which transfers the $U(1)$ eigenstates to the mass eigenstates and diagonalizes  the mass matrix via
\begin{equation}
\begin{pmatrix}
	\hat{S}_1 \\
	\hat{S}_2
\end{pmatrix}  =
U  \begin{pmatrix}
	S_1 \\
	S_2
\end{pmatrix} , ~~~
U^{\dagger} \cdot	\delta \hat{m}^2 \cdot U=\begin{pmatrix}
		\delta m^2_{11} & 0 \\
		0 &  \delta m^2_{22}
	\end{pmatrix} .
\end{equation}
Since the components proportional to identity matrix, $\delta m_{11}^2 \times I$, can be absorbed into the $U(1)$ conserving
mass term $m_S^2 \hat{S}^\dagger \hat{S}$, we can set $\delta m_{11}^2 = 0 $ without loss of generality. Because the Lagrangian $\mathcal{L}_\text{U(1)}$ is invariant under the rotation $U$, we obtain the Lagrangian in the mass eigenstates with DM mass and excited states mass respectively,
\begin{align}
	m_1^2 = m_S^2 , \quad m_2^2 = m_S^2 + \delta m_{22}^2 ,
\end{align}
where $\delta m_{22}^2 >0 $ is chosen, making $s_2$ the excited state.

The diagonalization of the mass matrix $\delta \hat{m}^2$ breaks the $U(1)$ global symmetry from random rotation to a special rotation angle $\theta$. Furthermore, the mass matrices $\delta \hat{m}^2$ and $\delta m^2$ are rank one, because $\delta m_{11}^2 = 0 $. It contributes a massive term $\delta m_{22}^2 s_2 s_2$ to the Lagrangian, while keeps $s_1$ mass unchanged.
In the aspect of global symmetry breaking, $s_2$ is similar to a radial mode, while $s_1$ is similar to the Goldstone mode after the symmetry breaking.
Actually, the special mass term $\delta m^2$ can be obtained by adding another complex scalar $\phi$ and assigning the global $U(1)$ charge $-2$ to $\phi$ and charge $1$ to $\hat S$.
Therefore, there is a new interaction term can be written as
\begin{align}
	\mu \phi \hat{S} \hat{S}  + h.c. ,
\end{align}
and the special rotation angle $\theta $ is actually
\begin{align}
\theta = \frac{1}{2}\text{arg}\left[ \mu \times \langle \phi  \rangle \right],
\end{align}
where $\langle \phi \rangle $ is the vev which explicitly breaks the global $U(1)$. After appropriately subtracting the identity component, one can obtain the required rank one mass matrix.

In principle, for the scalar quartic coupling $\hat{\lambda}_{ij} \hat{s}_i \hat{s}_j H^\dagger H$, it can exist
the $U(1)$ conserving component $|S|^2 H^\dagger H$, which can couple both $s_1 s_1$ and $s_2 s_2$ to the SM Higgs. However, it will make DM pair annihilation cross-section $\sigma_{11}$ comparable to the coannihilation partner annihilation cross-section $\sigma_{22}$ and the scenario comes back to the normal DM freeze-out. Therefore, we will omit the above parameter space and focus on the specific parameter space where $s_1 s_1$ does not couple to $H^\dagger H$.
Technically, it can be realized by adding the higher dimensional operator $\alpha \phi \hat{S} \hat{S} H^\dagger H + h.c.$ and \textit{require} that the complex phases of $\alpha$ and $\mu$ are the same.
In this case, the matrix $\hat{\lambda}$ is aligned with the special rotation angle $\phi$ and only $s_2 s_2$ couples to $H^\dagger H$. We emphasize that the $U(1)$ conserving component $|S|^2 H^\dagger H$ also respects the special rotation but is forbidden by hand. Therefore,
the above procedure actually picks up a specific interaction and leads to the parameter space which we are interested in. As a result, the following effective Lagrangian is our baseline model and in the mass eigenstates it reads,
\begin{align}
	\mathcal{L}_\text{eff} & =  \left(\partial_\mu S\right)^\dagger  \left(\partial^\mu S \right) - \frac{m_1^2}{2} s_1^2 - \frac{m_2^2}{2} s_2^2
	 - \lambda_{22} s_2^2 \left(H^\dagger H -\frac{v^2}{2}\right).
	 \label{eq:Leff}
\end{align}
This is the scalar iDM model to start with. It provides the mass splitting between dark matter ground state $s_1$
and  exited state $s_2$, and fulfills the requirement in Eq.~(\ref{eq:coannihilation-crosssection}).
In Eq.~(\ref{eq:Leff}),  there is no interaction between $s_1$ and $s_2$ yet, to provide the decay of $s_2$.  We will introduce two models for the decay of $s_2$ in the next subsection.

\subsection{The excited dark matter particle as long-lived particle}

Starting from the effective Lagrangian $\mathcal{L}_\text{eff}$, we have zero ground state annihilation $\sigma_{1 1} = 0$
and the coannihilation is dominant by $\sigma_{22}$. However, we should introduce a coupling between $s_1$ and $s_2$,
because $s_1$ has to be in thermal equilibrium with $s_2$ and the SM thermal bath. In addition, the coupling has to be small to make $s_2$ long-lived at the collider detector scale. We provide two models to achieve the above requirements.

\noindent $\bullet$ \textbf{Pure-Scalar model} (PS): we do not add new particles but slightly break the specialty of the angle $\theta$. Specifically, the mass matrix
$\delta \hat{m}^2$ and interaction matrix $\hat{\lambda}$ can commute with each other, $\left[\delta \hat{m}^2, ~\hat{\lambda} \right]=0$, thus they can be simultaneously diagonalized by a rotation matrix $U$. It means both matrices align to rotation angle $\theta$. Once the interaction is slightly misaligned to $\theta + \delta \theta$ with $ \delta \theta \ll \theta$, there are regenerated interactions between $s_1$--$s_2$ and $s_1$--$s_1$ itself
\begin{align}
	\lambda_{12} \approx - \delta \theta \times \lambda_{22}, \quad \lambda_{11} \approx  \delta \theta^2 \times \lambda_{22}.
\end{align}
Since $\delta \theta$ is very small, $	\lambda_{12}$ can lead to a slow decay of $s_2$. Because $\lambda_{11} $ is at the order of $\delta \theta^2$ which is negligible, thus the ground state annihilation contributes negligible cross section $\sigma_{1 1}$ comparing to the coannihilation. At leading order of $\delta \theta$, we denote the new contribution as the \textit{pure scalar model}
\begin{align}
	\mathcal{L}_{12}^{S} = - \lambda_{12} s_1s_2 \left(H^\dagger H -\frac{v^2}{2}\right) ,
\end{align}
with $\lambda_{12} \ll \lambda_{22}$.
Both annihilation process $s_1 s_2 \to \text{SM} \ \text{SM}$ and the decay width of $s_2$ are suppressed by $\lambda_{12}$. Moreover, the decay width of $s_2$ is additionally suppressed by small mass splitting $\Delta$ and small fermions mass in the Yukawa interaction. The $s_2$ decay width is approximately
\begin{equation}
	\Gamma (s_2 \to s_1 f \bar f) \simeq \frac{\lambda_{12}^2 m_f^2 m_2^3 \Delta^5}{240 \pi^3 m_h^4 }\times \theta(m_1\cdot\Delta-2m_f),
	\label{eq:s2decaySS}
\end{equation}
where the small mass $m_f$ is taken to be zero in the phase space integration.
For a typical electroweak mass, e.g. $m_2\sim 200$ GeV,  $s_2$ has a decay length (with all final decay states considered) of $\mathcal{O}( 10 ~{\rm cm}) $ for $\lambda_{12}\sim 10^{-3}$ and mass splitting $\Delta \sim 10\%$.
For massive $m_f$ comparable to mass splitting, one should numerically integrate the phase space to obtain the decay width.

\noindent $\bullet$ \textbf{Scalar-Vector model} (SV): we promote the global $U(1)$ to local $U(1)$, and keep the specialty of the rotation $\theta$ that $\left[\delta \hat{m}^2, ~\hat{\lambda} \right]=0$.
There is a massive dark photon $A'$,  from the $U(1)_D$ gauge field in the hidden sector, which can connect to SM particles via kinetic mixing term. The effective Lagrangian of DM and the Lagrangian of the dark photon are given below,
\begin{align}
      &   \mathcal{L}_\text{eff}  =  \left(D_\mu S\right)^\dagger  \left(D^\mu S \right)  - \frac{m_1^2}{2} s_1^2 - \frac{m_2^2}{2} s_2^2
	- \lambda_{22} s_2^2 \left(H^\dagger H -\frac{v^2}{2}\right),
	\label{eq:LeffSV}\\
	& \mathcal{L}_{A'}  = -\frac{1}{4} \hat{F}^{\prime \mu \nu}\hat{F}_{\mu \nu}^\prime -\frac{\epsilon}{2 c_w } \hat{F}^{\prime \mu \nu} B_{\mu \nu} + \frac{m_{A^\prime}^2}{2} \hat{A}^{\prime \mu} \hat{A}_{\mu}^\prime,
\end{align}
where $D_\mu=\partial_\mu-ig_D A'_\mu$ is the covariant derivative for the $U(1)_D$ interaction. $\hat{F}'$ is the field strength of $\hat{A}'$, $B$ is the field strength of the hypercharge field, $\epsilon$ is the strength of kinetic mixing, and $c_w$ is the cosine of the weak angle.

We can use a non-unitary matrix to rotate away the kinetic mixing terms and work in the mass eigenstates as follows  at leading order of $\epsilon$~\cite{Liu:2017lpo, Liu:2017zdh},
\begin{equation}
	\begin{pmatrix}
		\hat{Z}_\mu \\
		\hat{A}_\mu \\
		\hat{A}_\mu^\prime
	\end{pmatrix} =\begin{pmatrix}
		1 & 0 & \frac{m_{A^\prime}^2 t_w}{m_Z^2 -m_{A^\prime}^2}\epsilon\\
		0 & 1  & \epsilon \\
		\frac{m_{Z}^2 t_w}{ m_{A^\prime}^2 -m_Z^2}\epsilon  & 0 & 1
	\end{pmatrix}
	\begin{pmatrix}
		Z_\mu \\
		A_\mu \\
		A_\mu^\prime
	\end{pmatrix},
\end{equation}
where $A$, $Z$ and $A'$ are the photon, $Z$ boson in the SM and extra gauge boson from $U(1)_D$ in the mass eigenstate, while the expressions with a hat are for flavor basis.
The rotation matrix is expanded to $\mathcal O(\epsilon)$ and the mass $m_Z$ should not be too close to $m_{A^\prime}$.
The mixing among dark photon $A'$, $Z$ boson and massless photon gives rise to the coupling of $A'$ to the neutral current $J^\mu_Z$, electromagnetic current $J^\mu_{\text{em}}$ and dark current $J_D^\mu$. $Z$ boson also couples to the dark current $J^\mu_D$ due to the mixing. All of these interactions are suppressed by $\epsilon$.
Specifically, the interactions between mass eigenstate gauge bosons and currents are given in the following at leading order $\mathcal O(\epsilon)$,
\begin{align}
	\mathcal{L}_{\text{int}} &  = A_\mu e J_{\text{em}}^\mu  +
	Z_\mu \left(g J_Z^\mu - \epsilon g_D \frac{m_{Z}^2 t_w}{m_Z^2 -m_{A^\prime}^2} J_D^\mu \right)  +  A_\mu^\prime \left( g_D J_D^\mu + e \epsilon J_{\text{em}}^\mu + \epsilon g \frac{m_{A^\prime}^2 t_w}{m_Z^2 -m_{A^\prime}^2} J_Z^\mu  \right)  ,
\end{align}
where $J_D$ is the dark current for complex scalar $S$,
\begin{align}
	J_D^\mu = i\left( S^\dagger \partial^\mu S - S \partial^\mu S^\dagger \right)
	= s_2 \partial^\mu s_1 - s_1 \partial^\mu s_2,
\end{align}
which is invariant under global $U(1)_D$ rotation.

\begin{figure}[tbh]
	\centering
	\begin{align*}
		\begin{array}{cc}
			\includegraphics[scale=0.62]{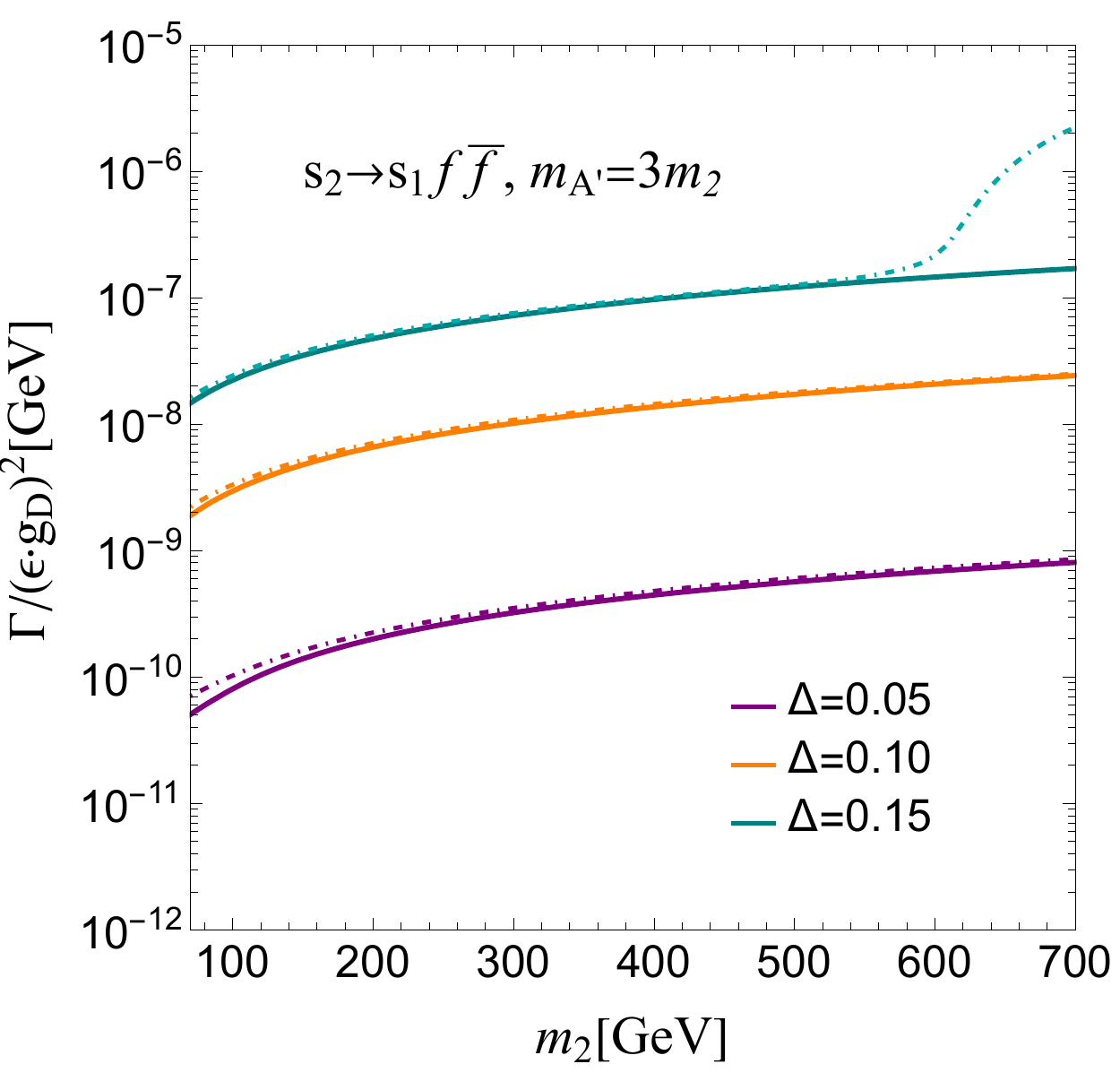}
			&
			\includegraphics[scale=0.625]{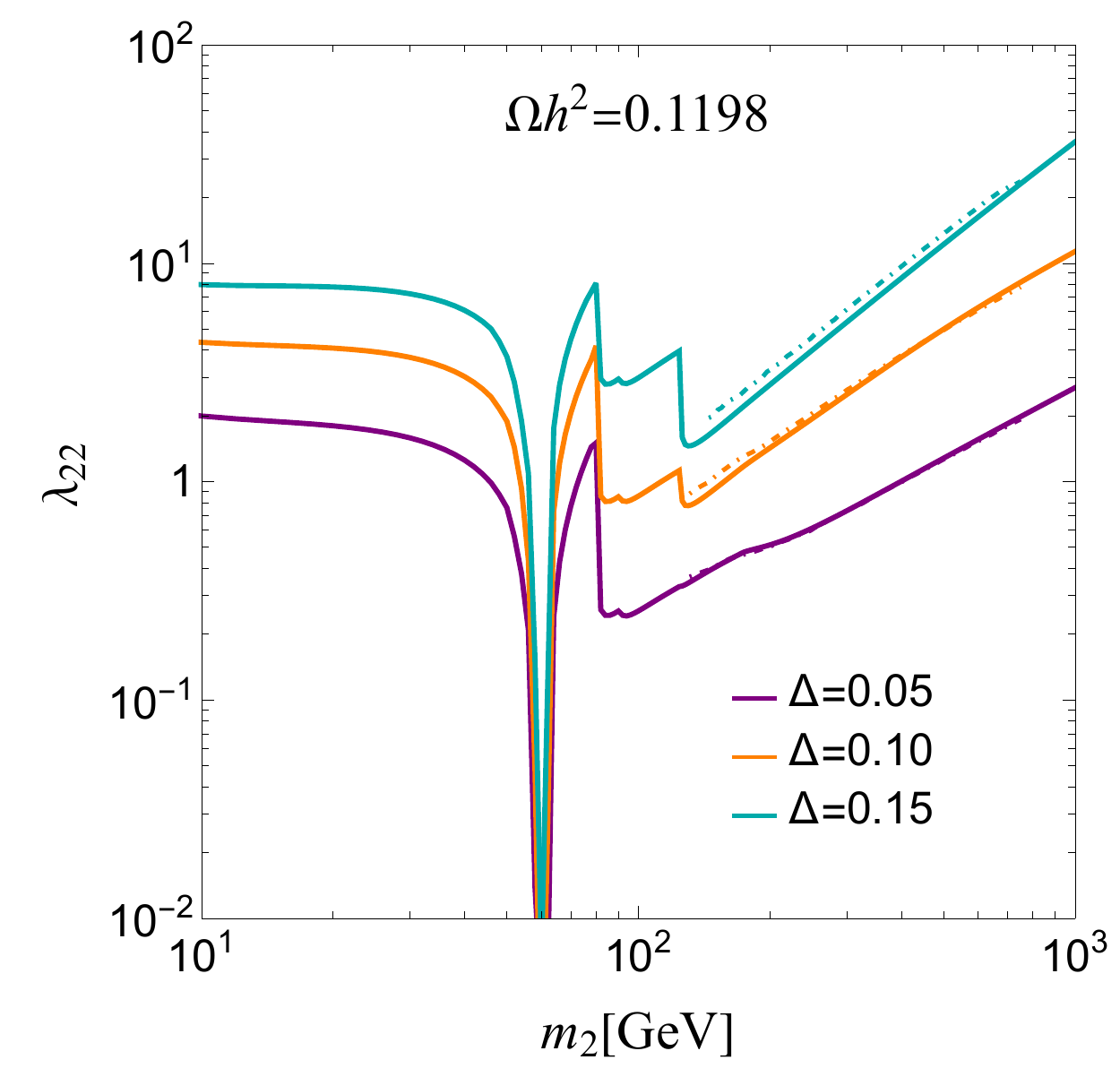}\\
			(a) & (b)
		\end{array}
	\end{align*}
	\caption{The left panel (a) shows the coannihilation partner $s_2$ total decay width ($\Delta=0.05,~0.10~ \rm{and}~ 0.15$) as a function of its mass $m_2$. The solid line and dashed line are our approximate and MadGraph results respectively. The dot-dashed cyan line has a difference at large mass for $\Delta = 0.15$, because the opening of a new channel $s_2 \to s_1 + Z$.
	The right panel (b) gives the parameter space of $m_2$ and $\lambda_{22}$ for the relic abundance. And the solid and dashed lines are numerical results and MadDM's results respectively.}
	\label{fig:RelDen}
\end{figure}

Moreover, $A'$ does not induce annihilation for $\sigma_{1 1}$ and $\sigma_{2 2}$. It only leads to the co-annihilation of $\sigma_{12}$ and the decay of $s_2$ as $s_2 \to s_1 f \bar{f}$.
Both $A'$ and $Z$ can mediate the decay $s_2\to s_1 f\bar f$, but the contribution from $Z$ boson involving $J_Z$ has an extra suppression factor of $( m_1\Delta)^2/m_Z^2 $ or $(m_1\Delta)^2/m_{A'}^2 $ comparing to the other contributions. This is because the $Z$ boson contribution will be almost canceled by the negative contribution from $A'$ when momentum transfer is small, e.g. $m_1\Delta \ll m_Z, ~ m_{A'}$~\cite{Liu:2017lpo}. As a result, the dominant contribution comes from the amplitude $\epsilon e g_D J_{\rm em }^\mu J_{D, \mu}/m_{A'}^2$ for heavy $A'$ mass.
In this case, the decay width of $s_2$ can be approximately written as
\begin{equation}
	\Gamma (s_2 \to s_1 f \bar{f}) \simeq Q_f^2 \frac{(\epsilon g_{D})^2 e^2 m_2^5 \Delta^5}{60 \pi^3 m_{A^\prime}^4 }\simeq Q_f^2 \left (\frac{\epsilon g_{D}}{10^{-3}}\right )^2\left(\frac{\Delta}{0.1}\right )^5\left(\frac{3m_2}{m_{A'}}\right)^4\frac{m_2}{100~\rm{GeV}}\cdot0.92~\rm{ns^{-1}},
	\label{eq:VecDec}
\end{equation}
where $Q_f$ is the electric charge of $f$ and $m_{A'}, m_2\gg m_f$. For the exact calculation and the plots, we use the numerical results from MadGraph for the final state phase space.

We show our approximate results and compare them to the MadGraph results in the left panel of  Fig.~\ref{fig:RelDen}. In this plot, we choose $m_{A'}= 3m_2$ and $\Delta=0.05, 0.10, 0.15$ as benchmark points. The decay width of $s_2$ increases with $m_{2}$, as shown in Eq.~(\ref{eq:VecDec}). Our approximate calculation and MadGraph results are consistent with each other when $m_1\Delta \ll m_Z$. After $m_1\Delta \gtrsim m_Z$, there will be a new decay channel $s_2\to s_1 Z$, leading to a significant increase of the decay width in the MadGraph results. Moreover, the cancellation of $J_Z$ contribution between $Z$ and $A'$ diagrams is not true anymore. For $\Delta =0.15$, this new channel opens around $600$ GeV, which is shown as the dark red line in the left panel of Fig.~\ref{fig:RelDen}; while for $\Delta =0.10$, this happens around $900$ GeV, which is outside of the plot range. For a long-lived $s_2 $ with a decay length around $10$ cm, the coupling $\epsilon g_D$ will be around $25~(4)\times 10^{-4}$ for $m_{2}= 300$ GeV and $\Delta=0.05 (0.10)$.
In the right panel of Fig.~\ref{fig:RelDen}, the parameter space of $m_2$ and $\lambda_{22}$ for the relic abundance are shown. The relic abundance is obtained with the help of coannihilating processes $s_2s_2\rightarrow \mathrm{SM \ SM}$. The corresponding $\lambda_{22}(m_2)$ will be used when generating the processes at LHC.

Lastly, there are four point vertex from $\mathcal{L}_{\text{eff}}$ in Eq.~(\ref{eq:LeffSV}), which is an exclusive feature for $U(1)_D$ charged scalar DM models,
\begin{equation}
	\frac{g_D^2}{2}(s_1^2+s_2^2) \left( A_\mu^\prime + \epsilon \frac{m_{Z}^2 t_w}{m_Z^2 -m_{A^\prime}^2} {Z}_\mu  \right)^2.
	\label{eq:4point-vertex}
\end{equation}
This term is again invariant under global $U(1)_D$ rotation, which will induces pair annihilation into $A'A', A'Z, ZZ$ gauge boson pair. In this work, we will set $m_{A'}= 3m_{2}$, that the only possible annihilation processes allowed by kinematics are $s_1 s_1(s_2s_2) \to ZZ$. However, such processes are suppressed by high power of $\epsilon g_D$ and the mass ratio $m_Z^2/m_{A'}^2$, which in total is $\mathcal{O}((\epsilon g_D m_Z/m_{A'})^4)$. Thus all the annihilation contributions from the four point interactions to $\sigma_{1 1,~22}$ can be neglected.

\section{Existing Constraints}
\label{sec:excon}

We will explore the potential of searching long-lived $s_2$ at the LHC experiments.
In this model, the DM obtains its right relic abundance dominantly through the coannihilation via the quartic interaction $\lambda_{22} s_2 s_2 H^\dagger H$. Therefore, the coupling $\lambda_{22}$ is sizable and we need to check the existing constraints from collider, direct and indirect experiments. Besides that, there are two more parameters $m_2$ and $\Delta$. For the coannihilation mechanism, the mass splitting $\Delta$ should not be too large and we take $0.05$, $0.10$ and $0.15$ as our benchmark points. For the mass parameter, we take it to be at electroweak scale and pay special attention for large mass $> 100$ GeV.
\\

\textbf{Relic Abundance}:
For pure-scalar and scalar-vector models, the excited state $s_2$ couples to SM Higgs via the quartic interaction $\lambda_{22} s_2 s_2 H^\dagger H$, which will lead to pair annihilation cross section for $s_2 s_2 \to {\rm SM} ~{\rm SM}$. Since the annihilation cross sections $\sigma_{1 1} $ and $\sigma_{12}$ are negligible,
the effective cross section $\sigma_{\text{eff}}$ is purely determined by $\lambda_{22}$ once the mass parameters are fixed.
\begin{figure}[h]
    \centering
    \includegraphics[scale=0.7]{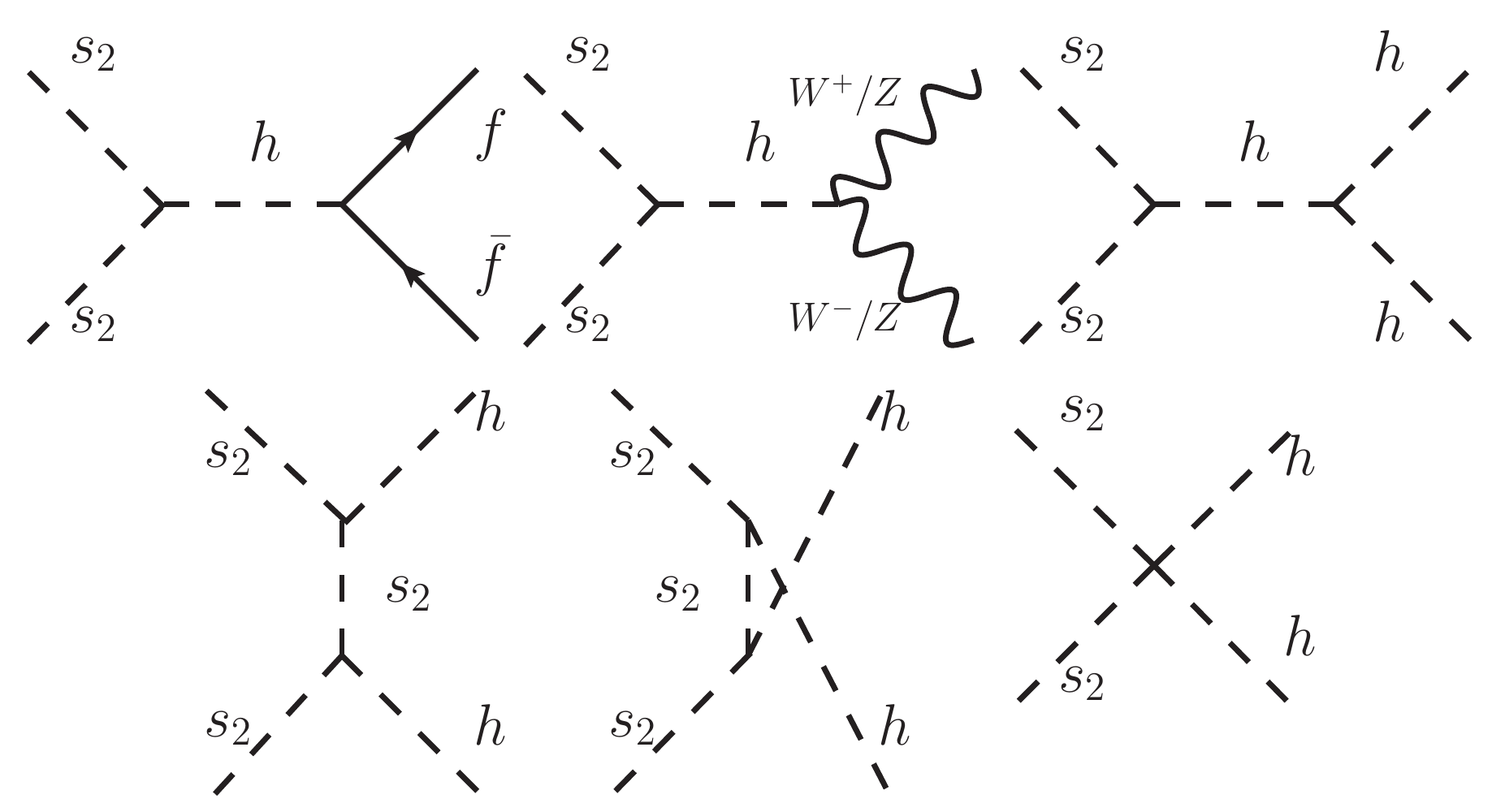}
    \caption{The Feynman diagrams for the annihilation $s_2s_2 \to {\rm SM} ~{\rm SM}$.}
    \label{fig:FeyDiag}
\end{figure}

The annihilation processes for  $s_2 s_2 \to {\rm SM} ~{\rm SM}$ include the final states $\bar{f}f$, $W^+ W^-$, $Z Z$ and $h h$, subjected to the kinematic constraints. The corresponding Feynman diagrams are given in Fig.~\ref{fig:FeyDiag}. The s-wave part of the cross sections $\langle \sigma v \rangle_{s}$ are given below,
\begin{align}
	\langle \sigma v \rangle_{s} =&\langle \sigma v \rangle_{f\bar{f}} +\langle \sigma v \rangle_{WW} +\langle \sigma v \rangle_{ZZ} +\langle \sigma v \rangle_{hh}, \\
	\langle \sigma v \rangle_{f\bar{f}} =&\frac{\lambda_{22}^2m_f^2(m_2^2-m_f^2)^{3/2}}{4\pi m_2^3(4m_2^2-m_h^2)^2},\\
	\label{annXSff}
	\langle \sigma v \rangle_{WW} =&\frac{\lambda_{22}^2(4m_2^2-4m_W^2m_2^2+3m_W^4)\sqrt{m_2^2-m_W^2}}{8\pi m_2^3(4m_2^2-m_h^2)^2},\\
	\langle \sigma v \rangle_{ZZ} =&\frac{\lambda_{22}^2(4m_2^2-4m_Z^2m_2^2+3m_Z^4)\sqrt{m_2^2-m_Z^2}}{16\pi m_2^3(4m_2^2-m_h^2)^2},\\
	\langle \sigma v \rangle_{hh} =&\frac{\lambda_{22}^2(\lambda_{22} v_h^2(4m_2^2-m_h^2)-4m_2^4+m_h^4)^2\sqrt{m_2^2-m_h^2}}{16\pi m_2^3\left(8m_2^4-6m_2^2m_h^2+m_h^4\right)^2}.
	\label{annXShh}
\end{align}
All of the annihilation cross sections are proportional to $\lambda_{22}^2$.
And the freeze-out temperature is determined by \cite{Griest:1990kh, Edsjo:1997bg, Bell:2013wua}
\begin{equation}\label{eq:xf}
    x_f = \ln{\frac{0.038 g_{\rm{eff}}m_{\rm{Pl}}m_1\langle \sigma_{\rm{eff}}v\rangle}{g_*^{1/2}x_f^{1/2}}}.
\end{equation}

And the relic abundance is
\begin{equation}\label{eq:omegah}
    \Omega h^2 = \frac{1.07\times10^9 }{g_*^{1/2}J(x_f)m_{\rm{Pl}}(\rm{GeV})},
\end{equation}
where $J(x_f)=\int_{x_f}^{\infty}\frac{{\langle \sigma_{\rm{eff}}v\rangle}}{x^2}dx$.
Together with Eq. (\ref{eq:sigma}), Eq.~(\ref{eq:xf}) and Eq.~(\ref{eq:omegah}), we can use numerical iteration to solve the freeze-out temperature $x_f$ and the coupling $\lambda_{22}$, which satisfies the DM relic abundance requirement. What's more, we find that the s-wave expansion of annihilation cross-section with small velocity might be invalid near the resonance region ($m_2\sim m_h/2$),
because there exists another small quantity $(4m_2^2 -m_h^2)/m_2^2$.
As a result, in order to avoid this effect, we consider its exact thermal average for $m_2<80\rm{GeV}$.

Besides, we also compare it to MadDM~\cite{Ambrogi:2018jqj} in the right panel of Fig.~\ref{fig:RelDen}. We can clearly see the analytic results are in agreement with MadDM's in the mass range $[100,700] \rm{~GeV}$ for  $\Delta=0.05$ and 0.10. While for $\Delta=0.15$ the MadDM's result is above numerical one, which shows the shortcomings of s-wave approximation. In order to understand the physics, we show the mass range from 10 GeV to 1000 GeV and $\Delta = 0.05, ~0.10, ~0.15$ respectively. It shows that the required $\lambda_{22}$ increases with $\Delta$ in general, due to the Boltzmann suppression factor $e^{-2 x \Delta}$. For light $s_2$ mass, e.g. $m_2 < m_h /2$, the required $\lambda_{22}$ is still larger than heavy $m_2$ region, because the opening channels are $f\bar{f}$ only which cross sections are suppressed by the small Yukawa couplings. There are dips around $m_{2} \sim m_h/2$ due to the SM Higgs resonance. The step features in the plot for large $m_2$ are originated from the opening of channels, $W^+ W^-$, $ZZ$, $hh$ and $t\bar{t}$ respectively. Besides, as shown in Fig. \ref{fig:RelDen}(b), for $\Delta=0.15$ when $m_2\gtrsim500\rm{GeV}$, the yukawa coupling $\lambda_{22}$ will
exceed $4\pi$ which violates the perturbation condition, so the red dashed lines in Fig. \ref{fig:VecRes} indicate this constraint.

In addition to the annihilation via the quartic interaction $\lambda_{22} s_2 s_2 H^\dagger H$, there are more annihilation channels for pure-scalar and scalar-vector model specifically.  For the pure-scalar case, there are also contributions from $s_1s_2\rightarrow h/s_2\rightarrow {\rm SM +SM}$ and $s_2s_2\rightarrow s_1\rightarrow {\rm SM +SM}$. However, these coannihilation cross sections are proportional to $\lambda_{12}^2$, which is tiny comparing to $\lambda_{22}^2$. Therefore we can safely ignore those contributions. For the scalar-vector model, there could be contributions from s-channel $s_1s_2\rightarrow A'/Z\rightarrow {\rm SM + SM}$ and t channel $s_2s_2/(s_1s_1)\rightarrow Z Z/(A'A')$, $s_1s_2\rightarrow h Z/A'$. The coannihilation cross sections of these processes are proportional to $\epsilon^2 g_D^2$, which is much smaller than $\lambda_{22}^2$. At the same time, dark photon mass $m_{A'}$ is set to be $m_{A^\prime}=3m_2$ to avoid annihilations to on-shell $A^\prime$. Thus we can ignore all these contributions to the relic abundance.
\\

\textbf{Thermalization}: The calculations above assume the equilibrium between $s_1$ and $s_2$ is achieved until freeze-out. The dominate relevant processes are up-scattering (down-scattering) with SM fermions, $ s_1+f \to s_2+f$. To achieve the equilibrium, we require $\Gamma > H $, where the rate $\Gamma$ defined as
\begin{equation}
    \Gamma(T) =\sum_{f} n^{eq}_f \langle \sigma_f v\rangle \gtrsim H,
    \label{eq:thermalize}
\end{equation}
where $ \sigma_f$ is the scattering cross section. The requirement can easily be satisfied at high temperature, but around freeze out, it require $\epsilon g_D \gtrsim 10^{-4} $. This constraint is shown in Fig. \ref{fig:VecRes}, where it cuts into the lower part of the LLP signal region.

Moreover, one can make a more careful treatment by solving the coupled Boltzmann equations, which is valid no matter the equilibrium maintained until freeze out or not. The equations are
\begin{equation}
	\left \{
	\begin{aligned}
		\frac{dY_1}{dx} &= -\frac{\lambda_f}{x^2}Y_f (Y_1 -\frac{Y_1^{eq}}{Y_2^{eq}} Y_2) + \gamma x (Y_2 -\frac{Y_2^{eq}}{Y_1^{eq}} Y_1), \\
		\frac{dY_2}{dx} &= -\frac{\lambda_{22}}{x^2} (Y_2^2 -Y_2^{eq2}) +\frac{\lambda_f}{x^2}Y_f (Y_1 -\frac{Y_1^{eq}}{Y_2^{eq}} Y_2) - \gamma x (Y_2 -\frac{Y_2^{eq}}{Y_1^{eq}} Y_1),
	\end{aligned}
	\right.
\end{equation}
where $\lambda=\frac{s(m_1)}{H(m_1)}\langle \sigma v\rangle$, $\gamma=\frac{\langle \Gamma_2 \rangle}{H(m_1)}$, and $\lambda_f$ is for up and down-scattering between the $s_1$ and $s_2$ while $\lambda_{22}$ is for $s_2$ annihilation into SM particles. 

\begin{figure}
	\centering
	\includegraphics[width=0.42\textwidth]{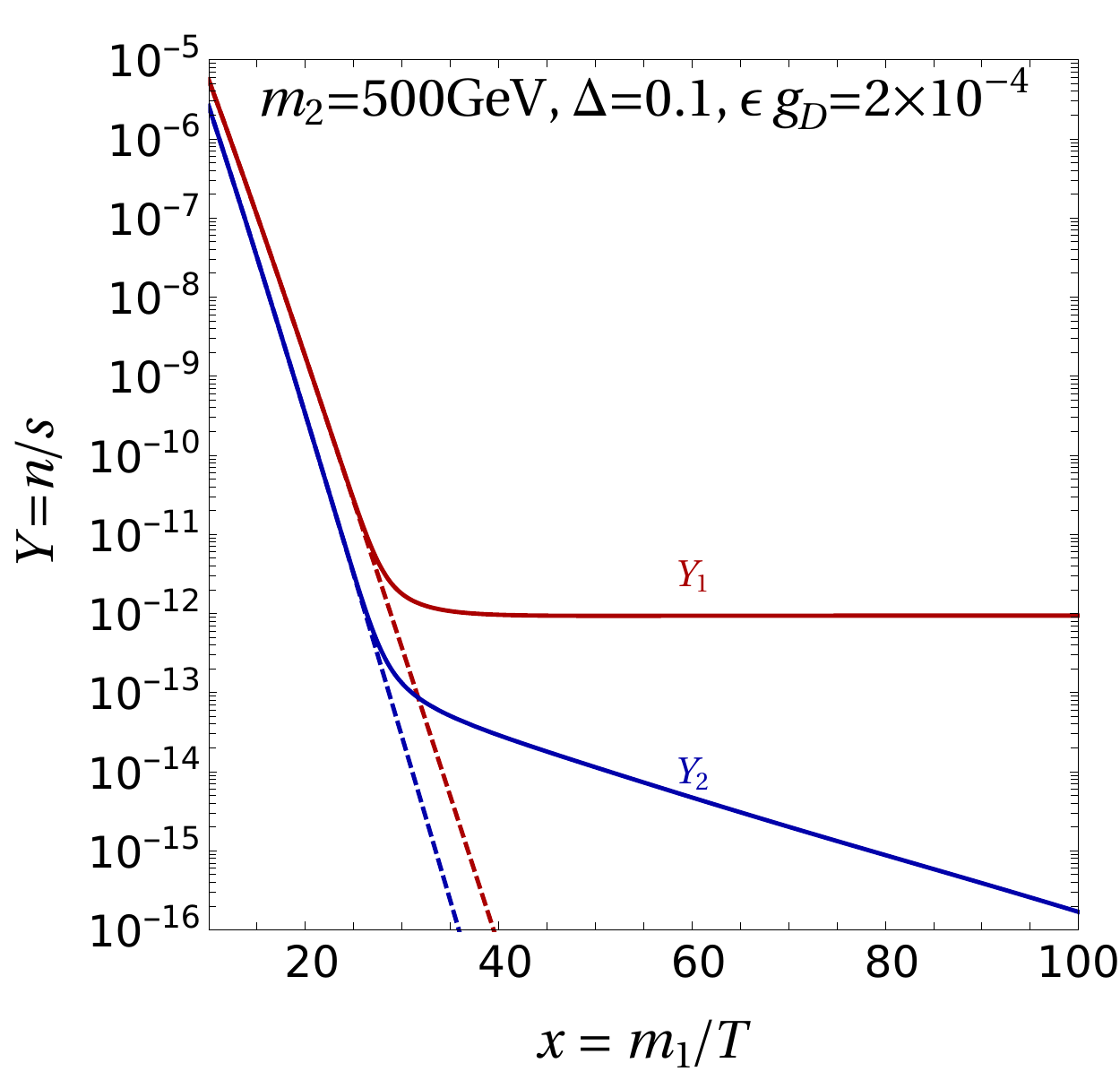}
	\includegraphics[width=0.42\textwidth]{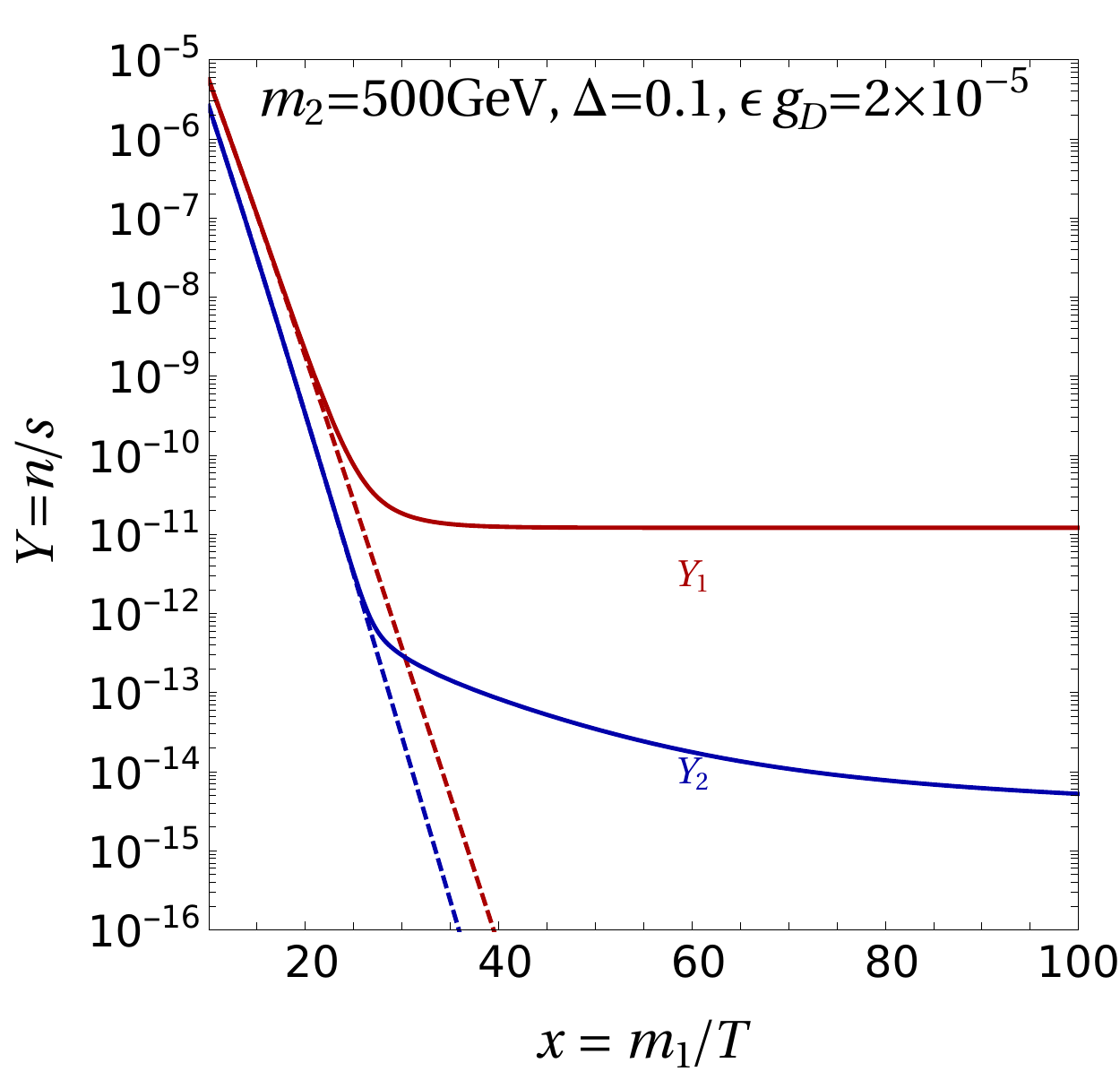}
	\caption{The solutions of coupled Boltzmann equations for two benchmark points. The left panel is for $m_2 =500 \text{GeV} , \  \Delta= 0.10 ,\ \epsilon g_D = 2 \times 10^{-4}$ while the right panel is for $m_2 =500 \text{GeV} , \  \Delta= 0.10 ,\ \epsilon g_D = 2 \times 10^{-5}$.}
	\label{fig:Bol}
\end{figure}

We have tested several benchmarks in our parameters and found that the results are in good agreement with our estimation using Eq.~(\ref{eq:thermalize}). In Fig.~\ref{fig:Bol}, we numerically solve the coupled Boltzmann equation and show the evolutions for the yield of $s_{1,2}$. We give two benchmark points with $\epsilon g_D $ above and below the thermalization estimation for $m_2 =500 \text{GeV} , \  \Delta= 0.1 $. In the case of $\epsilon g_D = 2 \times 10^{-4}$, we find it can satisfy the DM relic abundance $\Omega h^2 = 0.117$. However, in the other case of $\epsilon g_D = 2 \times 10^{-5}$, we find that DM relic abundance is too large, $\Omega h^2 = 1.509$, because DM freeze-out happens too early. 
\\

\textbf{Indirect Detection}:
In our model, the only significant annihilation to SM particles are from $s_2+s_2$. However, the life-time of $s_2$ is quite short comparing with the Hubble, thus $s_2$ already decays before CMB. Therefore, it does not inject energy to the thermal plasma during CMB era or after. While for $s_1$, it can have the annihilation channel $s_1 s_1 \to h h$ via t-channel $s_2$, but is suppressed by small $\lambda_{12}^2 \sim 10^{-6}$ if requiring $c \tau_{s_2} \sim 10 ~{\rm cm}$. For the vector-scalar case, there could be annihilation channel $s_1s_1 \to Z Z$ via t-channel $s_2$ or four point vertex in Eq.~(\ref{eq:4point-vertex}), but is suppressed by $\epsilon^2$.
Therefore, due to the absence of $s_2$ in the late universe and the small annihilation cross section of $s_1 s_1$, the indirect detection constraints can not restrain the scalar iDM model.
\\

 \textbf{Direct Detection}:
The DM $s_1$ does not couple to SM particles directly, so  the tree-level contribution in dark matter-nucleus/electron elastic scattering is missing. It is a result from the condition $\sigma_{1 1} \simeq 0$. When going to the full models with $s_2$ decay, the direct detection cross section should be considered with the presence of $s_2$. In the pure-scalar case, the coupling $\lambda_{12}$ will induce loop-level scattering cross section~\cite{Casas:2017jjg}. The spin independent direct detection cross section will be suppressed by $\lambda_{12}^4/(16\pi^2)^2$, which is too small to be constrained. On the other hand, there  could be inelastic scattering process for direct detection $s_1 N\to s_2 N$ induced by $\lambda_{12}$. But our typical mass difference is $m_1\cdot \Delta > 1$ GeV, which is significantly much larger than the kinetic energy of non-relativistic $s_1$. Thus, the inelastic scattering is forbidden by the kinematics. For scalar-vector model, there are 1-loop diagram contributions for elastic scattering, via a box diagram mediated by $s_2$ and a triangle diagram from Eq.~(\ref{eq:4point-vertex}) which is special for scalar DM. Such contributions are proportional to $\epsilon^4$ and further suppressed by high powers of $m^2_Z/m_{A'}^2$ and loop factors, thus direct detection experiments does not constrain our  parameter space~\cite{Izaguirre:2015zva, Berlin:2018jbm}.
\\

\textbf{LHC and Electroweak Precision Test}:
The coannihilation mechanism requires a large coupling to SM particles, which is realized by the quartic scalar coupling $\lambda_{22}$. Through this interaction, the LHC can produce $s_2$ pair through the Higgs mediated process $pp \to s_2 s_2$, followed by the $s_2$ decay $s_2 \to s_1 f \bar{f}$. Since the mass difference between $s_2$ and $s_1$ is about $\sim 10\%$, the fermions in the final states are quite soft to detect. However, with an extra energetic initial radiation jet, the process $pp \to j+s_2 s_2$ has the same feature as the  mono-jet plus missing energy. Therefore, it can be constrained by mono-jet searches at LHC \cite{Aad:2021egl,CMS:2021far}. Our signal cross section without cut is less than 100 fb after fixing $\lambda_{22}$ by the relic abundance, for $m_2\in [70, 700]$ GeV. The LHC constraint on the cross section is $\sigma  A  \epsilon < 736$ fb with some basic cuts on $p_T$ and $\slashed{E}_T$ and acceptance efficiency included, therefore the model we consider is safe from the mono-jet searches.

For the scalar-vector model, there are additional constraints because the dark photon $A'$ couples to the electromagnetic current with the coupling strength $\epsilon e$. One important constraint comes from the dilepton resonance search \cite{Aad:2019fac, Sirunyan:2021khd}, which sets limit on $\sigma(pp\to A'){\rm BR}(A'\to \ell^+\ell^-)$. Such cross section is proportional to $\epsilon^2$,
however the branching ratio ${\rm BR}(A'\to \ell^+\ell^-)$ depends on both $\epsilon$ and $g_D$ due to the DM decay channel $A' \to s_1 s_2$.
In this study, we fix $m_{A'}=3 m_2$ and in Fig.~\ref{fig:VecRes} we choose
$g_D=0.1$ as a benchmark point. In this case, $\lambda_{22} \gg g_D, \epsilon$, so that the coannihilation are dominated by $s_2 s_2 \to {\rm SM}~ {\rm SM}$ and the other coannihillation processes are suppressed by small $g_D$ or $\epsilon$.
We find that the constraint from dilepton searches at LHC requires $\epsilon \lesssim 0.03 \text{--} 0.1$ for $m_{A'} \in \left[100,~600\right]$ GeV respectively, as shown in gray shaded region in Fig.~\ref{fig:VecRes}.
Another relevant constraint for scalar-vector model comes from the electroweak precision test (EWPT) \cite{Curtin:2014cca}, because the mixing between the dark photon and the $Z$ gauge boson. The kinematic mixing from $A'$ can shift $Z$ boson mass and its couplings to SM fermions, thus affects the global fitting of the electroweak observable. For our setup, the EWPT constraint is weaker than the dilepton resonance searches.  We plot the relevant constraints in Fig.~\ref{fig:VecRes}, which are complementary to the sensitive region from the LLP searches.

\section{Long-lived particle signatures of the excited dark matter particle }
\label{sec:LLP}

\subsection{The production and decay of the long-lived particle}
\begin{figure}[htbp]
	\centering
	\begin{equation*}
		\begin{array}{cc}
			\includegraphics[width=0.5\linewidth]{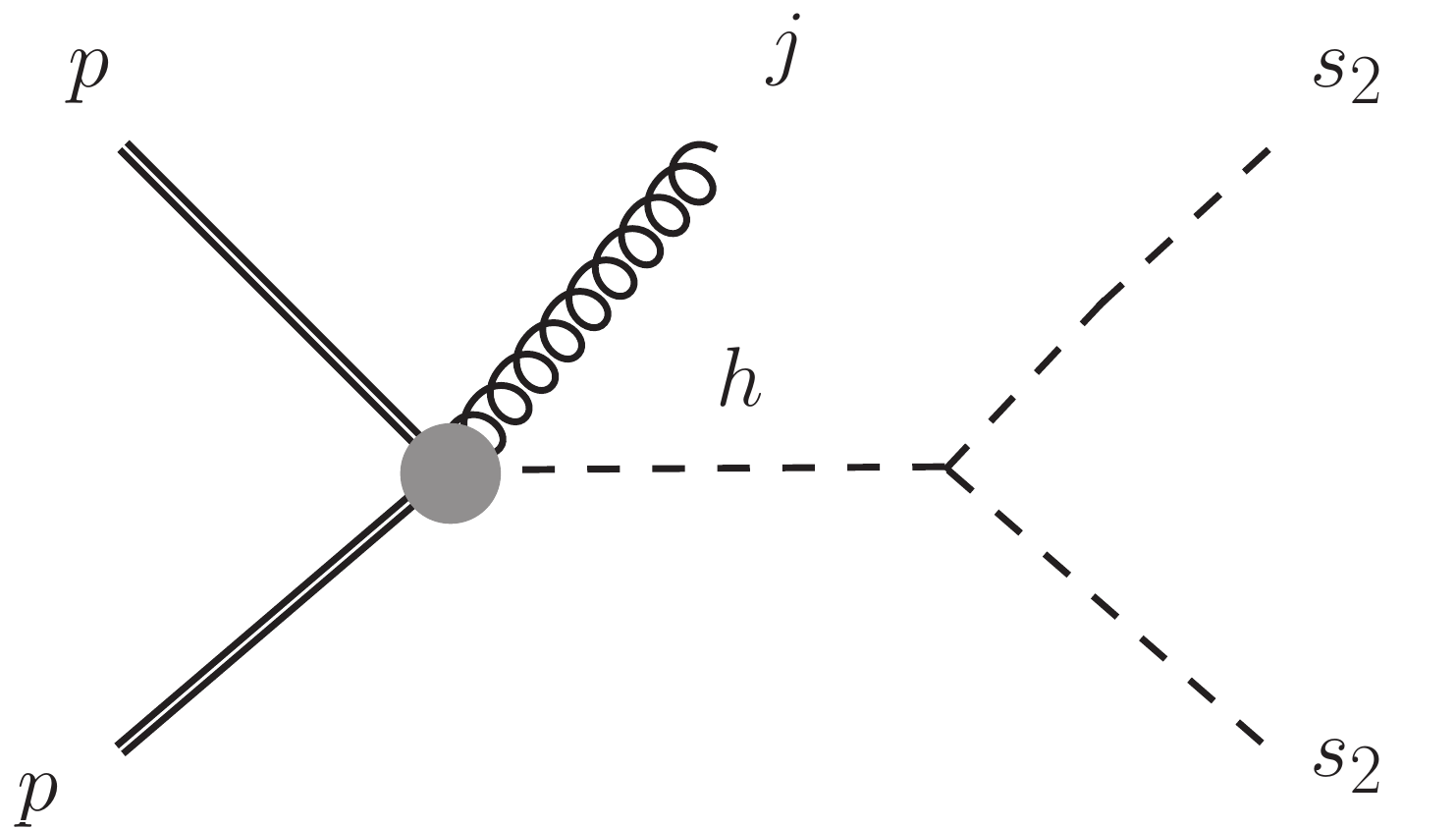}
			&
			\includegraphics[width=0.5\linewidth]{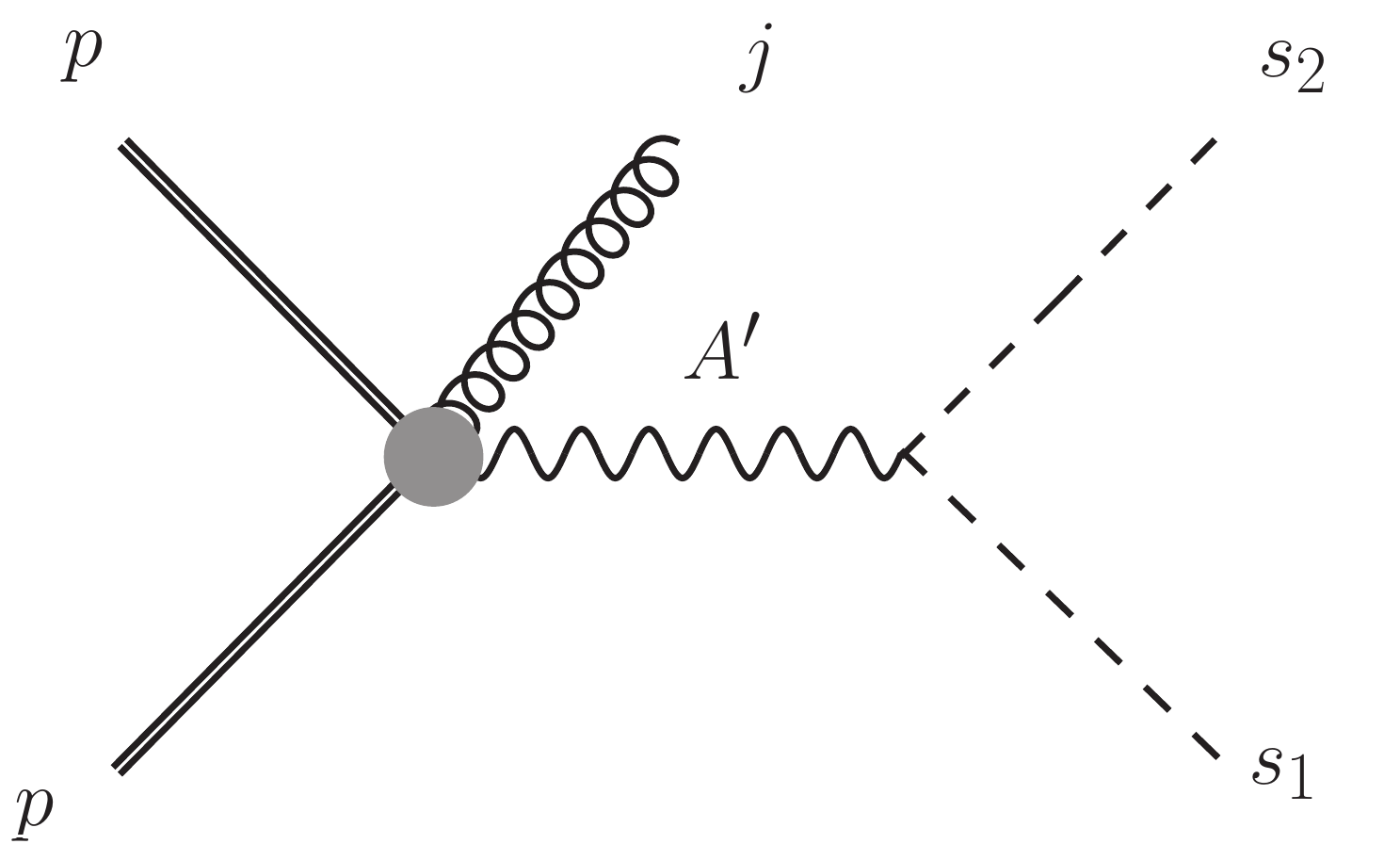}\\
			(a) & (b)
		\end{array}
	\end{equation*}
	\caption{The Feynman diagrams for LHC productions of $s_2$ in the scalar-vector model.}
	\label{fig:vec}
\end{figure}

We are interested in the dark sector particles with mass $m_{1,2} \gtrsim \mathcal O(100)\text{GeV}$,  therefore LHC is the most appropriate experiment to look for it.  In this section, we discuss the probes of coannihilating DM and its partner at the future high-luminosity LHC (HL-LHC), with the integrated luminosity $\mathcal{L}=3~\mathrm{ab^{-1}}$. For the pure-scalar and scalar-vector models, one can produce the excited states $s_2$ through Higgs portal or dark photon with an initial state radiation jet, namely
\begin{align}
	p p \to jh^{*} \to j s_2 s_2 , ~~ p p \to j A'\to j s_2 s_1.
	\label{eq:production}
\end{align}
The Feynman diagrams are listed in Fig.~\ref{fig:vec}.

The $s_2 s_2$ is produced via s-channel off-shell SM Higgs in both two models, while the $s_2 s_1$ production on the right of Eq.~(\ref{eq:production}) is specific to the scalar-vector model for heavy $s_{1,2}$, because the $A'$ is heavy enough to decay to $s_2 s_1$ but SM Higgs can not decay to $s_2 s_1$. The first process cross section is only determined by the $s_2$ mass after fix $\lambda_{22}$ via the dark matter relic abundance. While for the second process, the cross section depends on $\epsilon$ and $g_D$ together with the $m_2$, even after we fix $A'$ mass as $m_{A'}=3m_2$. In our study, we focus on the case $\lambda_{22} \gg \epsilon, g_D$, therefore the first one will be the dominant process to search at HL-LHC.  As a coannihilation partner, $s_2$ is unstable and subsequently decays to $s_1$ and SM particles as
\begin{align}
	s_2 \to s_1 + jj, \quad s_2 \to s_1 + \ell^+ \ell^- .
\end{align}
The former one happens for both pure-scalar and scalar-vector models, and the second one can have a significant
branching ratio for scalar-vector model only because of the small lepton mass suppression in Yukawa coupling in pure-scalar model.
The leptons are much easier to search at LHC comparing to jets, especially for soft objects. As a result, in this study we will focus on the scalar-vector model and the leptonic decay $s_2 \to s_1 + \ell^+ \ell^-$.

\subsection{The generic features of the LLPs }

For the neutral LLP $s_2$, its decay can be spatially displaced and also time delayed, depending on its mass and mass splitting. Inside the detector, the decay products of $s_2$ can be reconstructed as a displaced vertex, which is spatially separated from the interaction point. Therefore, it is different from most of the
SM backgrounds which are prompt and can be used to suppress the SM background.
Regarding the time delay, it comes from the slow movement of the heavy $s_2$, which results a time delayed arrival at the detectors.
In the future upgrade of the HL-LHC, the timing layers are deployed to suppress the pile-up events and more precise measurements for location, momentum and energy of the particles. For example, CMS is working on the minimum ionizing particle (MIP) timing detector~\cite{CERN-LHCC-2017-027, Contardo:2020886}, ATLAS is working on
the High Granularity Timing Detector \cite{Allaire:2018bof} and LHCb has the
similar precision timing upgrades in the future \cite{LHCb:2018hne}.
For SM particles, especially the mesons and leptons, they are moving at the speed of light. The heavier objects in the SM decay instantly into the light particles, therefore they also have no time lag and their signals arrive at the detector very fast. As a result, the heavy $s_2$ can significantly lag behind the SM process in time.
A quantitative description of the time difference is given as~\cite{Liu_2019}
\begin{equation}
	\Delta t_f=L_{s_2}/\beta_{s_2}+L_{f}/\beta_{f}-L_{\mathrm{SM}}/\beta_{\mathrm{SM}},
	\label{eq:timedelcay}
\end{equation}
for the decay $s_2 \to s_1 + \bar{f}f$, where $\beta$ and $L$ denotes the velocity and the moving distance of each particle, and ${\rm SM}$  denotes a trajectory connecting interaction point and the arrival point at the detector via a SM particle.
For simplicity, the trajectories of $s_2$ and decay products are assumed to be straight lines, and $\beta_{f}\simeq\beta_{\mathrm{SM}}\simeq1$ are adopted.
For $b$ quark or $\tau$ lepton, they are heavy but decay fairly quickly into light leptons, mesons or hadrons, which are again ultra-relativistic. Therefore, the above assumptions are viable.

Regarding the signal trigger, we always require an initial state radiation jet accompanied with the signal, which can time stamp the primary vertex~\cite{Liu_2019}. A hard initial state radiation jet with $p_T^j > 120$ GeV can also trigger the signal event with Jet$+$MET tagger~\cite{CMS:2014jvv, CMS:2019ctu}.
There are other triggers which can help loosen the requirements on the hard leading jet. For example, people have discussed using the displaced track information to implement the L1 hardware trigger, and the requirement on the track $p_T$ can be as low as $2$ GeV~\cite{Bartz:2017nlo, Tomalin:2017hts, CMS-PAS-FTR-18-018, Gershtein:2017tsv, Martensson:2019sfa, Gershtein:2019dhy, Ryd:2020ear, Gershtein:2020mwi}. The delayed photon and jet are studied in Ref.~\cite{ATLAS:2014kbb, Liu_2019, CMS:2019qjk} to set limits for LLPs. Using delayed objects for trigger is under discussion and development~\cite{CERN-LHCC-2017-027}.
In the ATLAS experiment, one can also use the Muon Spectrometer Region of Interest method to trigger the displaced events~\cite{ATLAS:2015xit}. In summary, there are many ways to improve the triggers for the LLP signal. As a result, a trigger with a hard initial jet radiation is quite conservative and could be further improved.
With the presence of leptons, the trigger becomes even more easier comparing with pure hadronic final states.
The specific triggers, signal cuts and the background estimates will be addressed in the later subsections.

Besides the ATLAS and CMS experiments, there are also dedicated experiments or future plans for LLPs, such as MATHUSLA (MAssive Timing Hodoscope for Ultra-Stable neutraL pArticles)~\cite{Chou:2016lxi, Curtin:2018mvb}, FASER \cite{Kling:2018wct,Feng:2017uoz}, CODEX-b \cite{Gligorov:2017nwh}. We consider all of them and find that the MATHUSLA experiment is much better than FASER and CODEX-b due to the specific model and the parameter space we are interested in.
We stress that the work will focus on the scalar-vector model in the LLP study. The signature of pure-scalar model includes soft jets, which trigger and QCD background are very challenging. Some track based strategies may reduced background\cite{Liu:2020vur, Hook:2019qoh}, but we will leave it for the future work.

\subsection{The scalar-vector model at LHC}

In the scalar-vector model, there are two production channels for $s_2$, which are shown in Fig.~\ref{fig:vec}. The left panel is realized via off-shell Higgs boson, and the right panel is realized via on-shell $A'$. As shown in the right panel of Fig.~\ref{fig:RelDen}, one needs $\lambda_{22}\sim \mathcal{O}(1)$ to realize the right dark matter relic abundance. Since the $\epsilon$ and $g_D$ are much smaller than $\lambda_{22}$, the main production channel of $s_2$ in LHC is by exchanging off-shell Higgs and its cross section is proportional to $\lambda_{22}^2$. The other production channel $p p \to j A'$ is proportional to $\epsilon^2$. When $\epsilon$ is large enough, the on-shell production of $A'$, followed by $A'\to s_1 s_2$ decay is considered in our calculation. In this work, we fix $g_D =0.1$ and $m_{A' } =  3 m_2$ as our benchmark point to reduce the parameters.

The excited state $s_2$ couples to $s_1$ mainly through dark photon $A'$. Since we assume heavy $A'$, then $s_2$ will only decay to $s_1 f \bar{f}$ via off-shell $A'$. Because $A'$ couples to all the SM electromagnetic current via the kinetic mixing, for a reasonable consideration we can have $2m_b < m_1\cdot\Delta < 2m_t$, with $b,t$ denoting bottom and top quarks. The total width of $s_2$ is
\begin{align}
\Gamma_{s_2} \simeq \frac{(\epsilon g_{D} e)^2 m_2^5 \Delta^5}{9 \pi^3 m_{A^\prime}^4 },
\label{eq:30}
\end{align}
which is shown in the left panel of Fig.~\ref{fig:RelDen}. The signals we consider for scalar-vector model are
   \begin{align}
 pp&\rightarrow j s_2 s_2~(j s_2 s_1), \quad  s_2  \to s_1 \ell^+\ell^- .
\end{align}
We take the inclusive strategy that at least one of $s_2$ decays to leptons in the detector.
The branching ratio of $s_2\to s_1{ ~\ell^+ ~ \ell^-}$ can be estimated as $3/10$ by counting the degrees of freedom of the particles, where $\ell = e, ~ \mu$.
Another important physical parameter is the lifetime of $s_2$, $\tau=1/\Gamma$, for the LLP searches at HL-LHC. The last important free parameter is the mass difference $\Delta$, which is important for triggering the signal via the leptons. In summary, there are only three free physical parameters, after we assume $g_D=0.1$, $m_{A'}=3 m_{2}$ and fix $\lambda_{22}$ by relic abundance, which are
\begin{align}
\left\{ m_2, \Delta, \epsilon \right\}.
\end{align}

There are many strategies to look for LLPs together with different triggers~\cite{Alimena:2021mdu}.
Since the $s_2$'s decay products contain leptons, it is easier to trigger. For example, in CMS Run-2, the scouting technique has been used to select two muons events with $p_T$ as low as $3$ GeV~\cite{Alimena:2021mdu}.
One search strategy relies on the presence of displaced muons, denoted as \textit{displaced muon-jet} (DMJ) \cite{Izaguirre:2015zva}, and worked conservatively with the Jet$+$MET trigger~\cite{CMS:2014jvv, CMS:2019ctu}. Therefore, the detailed cuts are~\cite{Izaguirre:2015zva, Berlin:2018jbm},
\begin{align}
	\text{DMJ}: ~p_T^j > 120~\text{GeV}, ~~p_T^{\mu} > 5~\text{GeV}, ~~ r_{s_2} < 30 ~ \text{cm},
	~~d_0^\mu > 1 ~\text{mm},
\end{align}
where $r_{s_2}$ is a radial displacement of the $s_2$ decay vertex and $d_0 $ is  transverse impact parameter.  The condition $r_{s_2} < 30 ~ \text{cm}$ guarantees the $s_2$ decay leaves tracks in the tracking system. The backgrounds can be reduced to a negligible level after the above cuts~\cite{Izaguirre:2015zva}.

Another possible strategy utilizes the time delay of heavy LLP, and the leptons are not specified to muons~\cite{Berlin:2018jbm}. Specifically, the cuts are taken as
\begin{align}
{\rm Timing:~~ }& p_{T}^j>120~\mathrm{GeV}~ (30~\mathrm{GeV}), \ \
p_{T}^\ell>3~\mathrm{GeV},  \ \
|\eta|<2.4, \nonumber \\
& \Delta t_\ell>0.3~\mathrm{ns},   \ \
5~{\rm cm}<r_{s_2}<1.17~\text{m} ,\ \
z_{s_2}<3.04~\mathrm{m},
\label{eq:siglltimingtrigger}
\end{align}
where $\eta$ is the pseudo-rapidity for the jet and leptons. The time delay for leptons $\Delta t_\ell$ are used to suppress the SM background. The radius and longitudinal location of the decay vertex, $r_{s_2}$ and $z_{s_2}$, have to be within the CMS MIP timing detector to ensure the hits on the timing layer.
For the initial state radiation, the $p_T^j$ cut has two choices. One is conservative,  $p_{T}^j>120~\mathrm{GeV}$, which is used by the conventional Jet$+$MET trigger. On the other hand, one can also be optimistic with timing information and the presence of the leptons, that a lower threshold $p_{T}^j>30~\mathrm{GeV}$ is possible in the near future.
The backgrounds can be sufficiently suppressed with the above cuts, therefore the SM backgrounds are taken to be zero~\cite{Berlin:2018jbm, Liu_2019}.

Aside from LHC detectors, MATHUSLA is a proposed LLP detector at CERN, located on the surface. The main detector is 20 meters tall and $200{\rm m} \times 200{\rm m}$ in area. MATHUSLA is shielded by $\sim 100$ m of rock to keep out of QCD backgrounds. The bottom and side of MATHUSLA are covered with scintillator to veto incoming charged particles, such as high energy muons and cosmic rays. In conclusion, the LLP search at MATHUSLA can be assumed to be background free. In order to consider the sensitivity on $s_2$ search at MATHUSLA, we require $s_2$ to decay inside its decay volume,
\begin{align}
	{\rm MATHUSLA}:   100~{\rm m}< x_{s_2} < 120\text{m} , -100 {\rm m}<y_{s_2}< 100\text{m} ,  100~{\rm m}<z_{s_2}< 300\text{m}.
\label{eq:sigJJcuts}
\end{align}
One considers the signals as charged tracks with energy deposition of more than 600 MeV, following the discussion  in Ref.~\cite{Curtin:2018mvb}.

\begin{figure}[ht]
	\centering
	\includegraphics[width=0.32\textwidth]{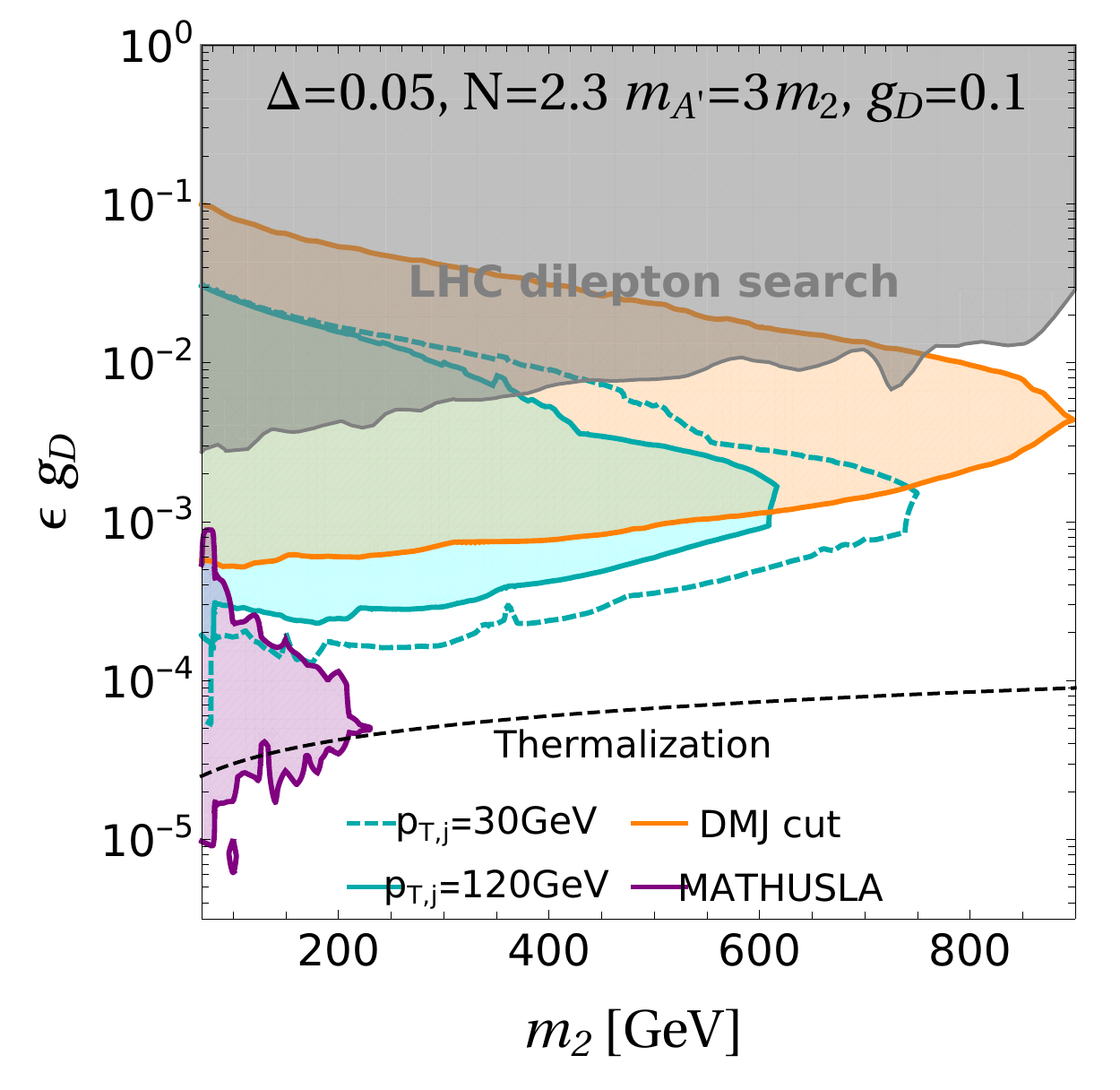}
	\includegraphics[width=0.32\textwidth]{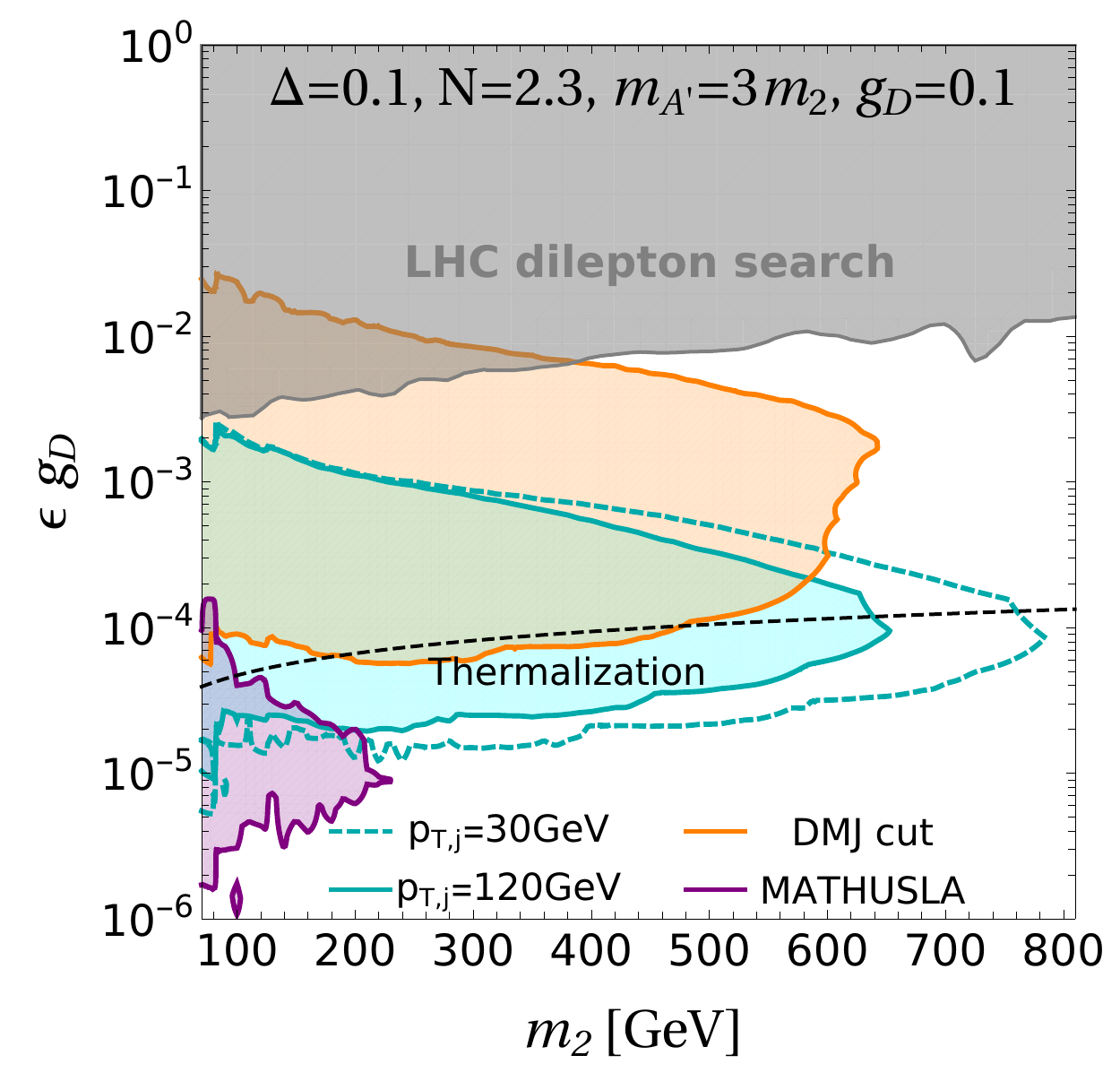}
	\includegraphics[width=0.32\textwidth]{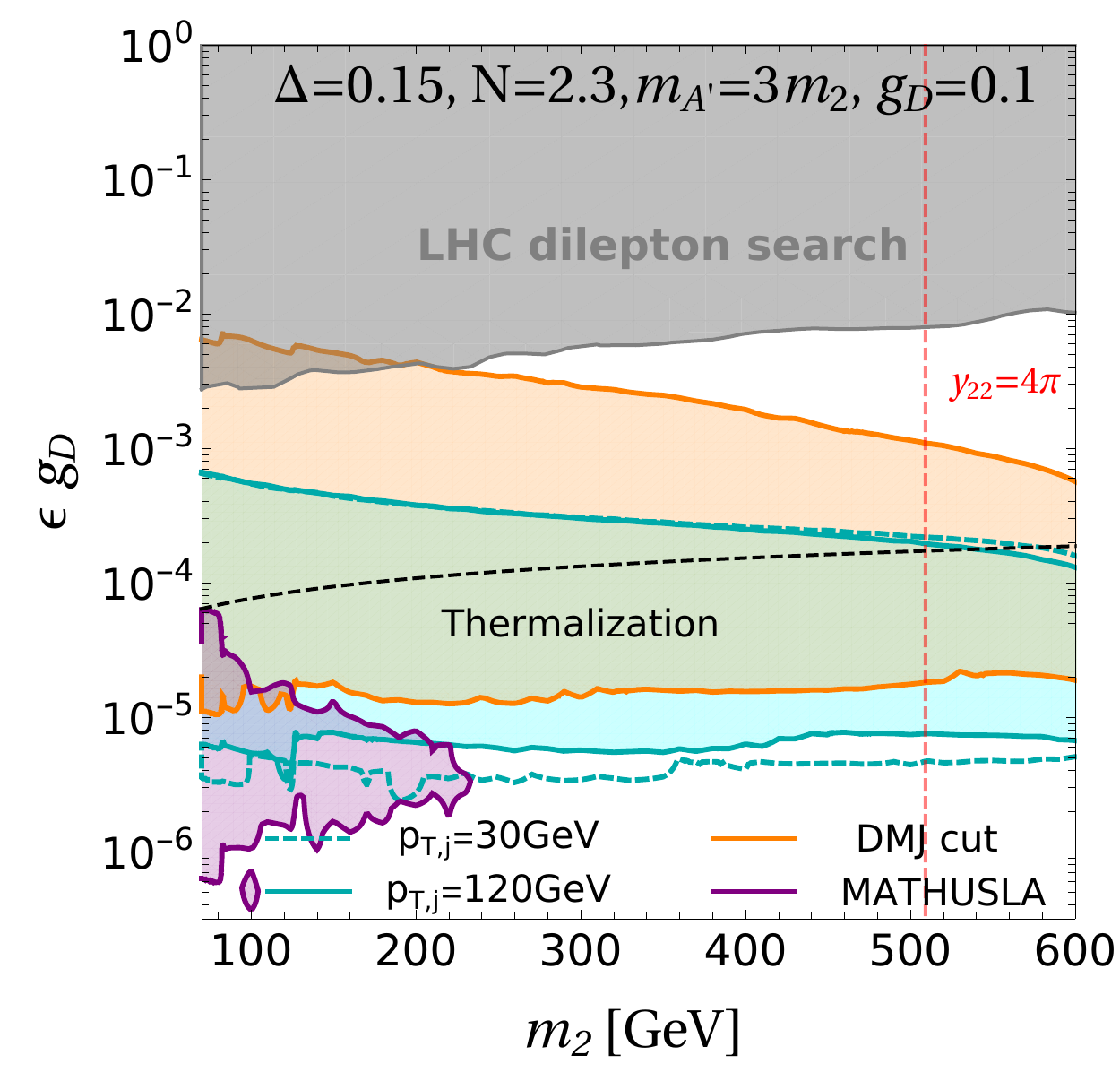}
	\includegraphics[width=0.32\textwidth]{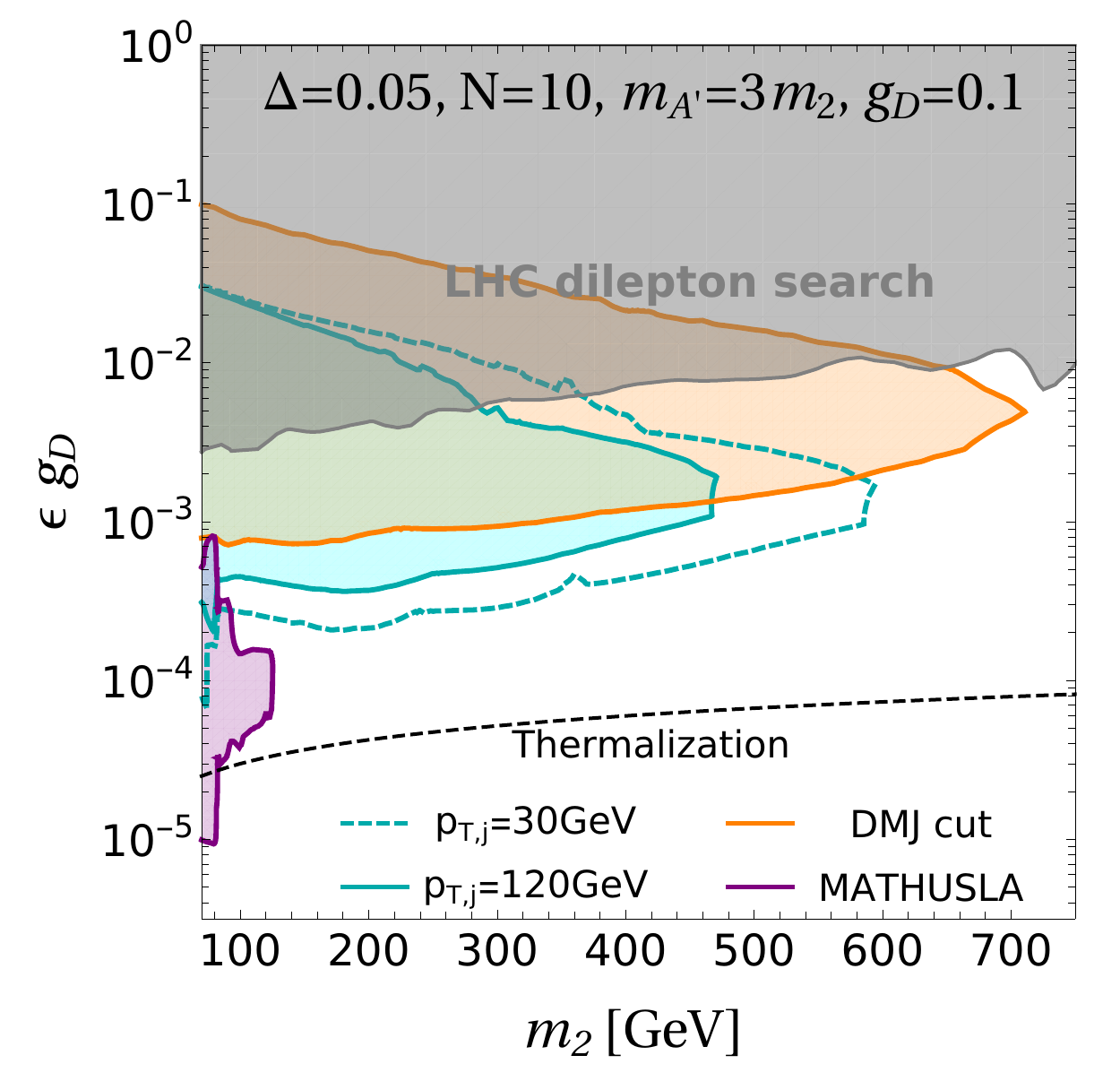}
	\includegraphics[width=0.32\textwidth]{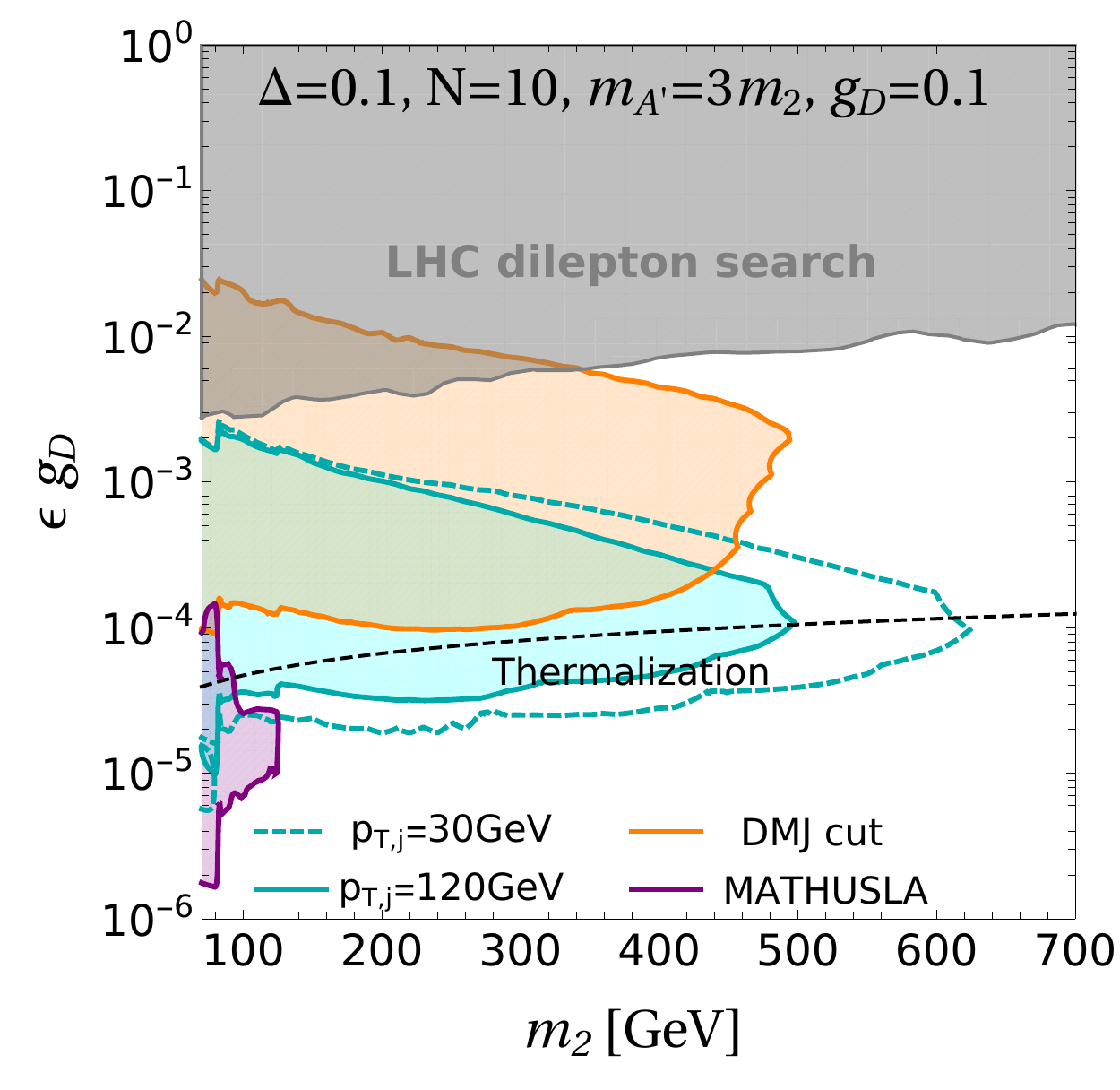}
	\includegraphics[width=0.32\textwidth]{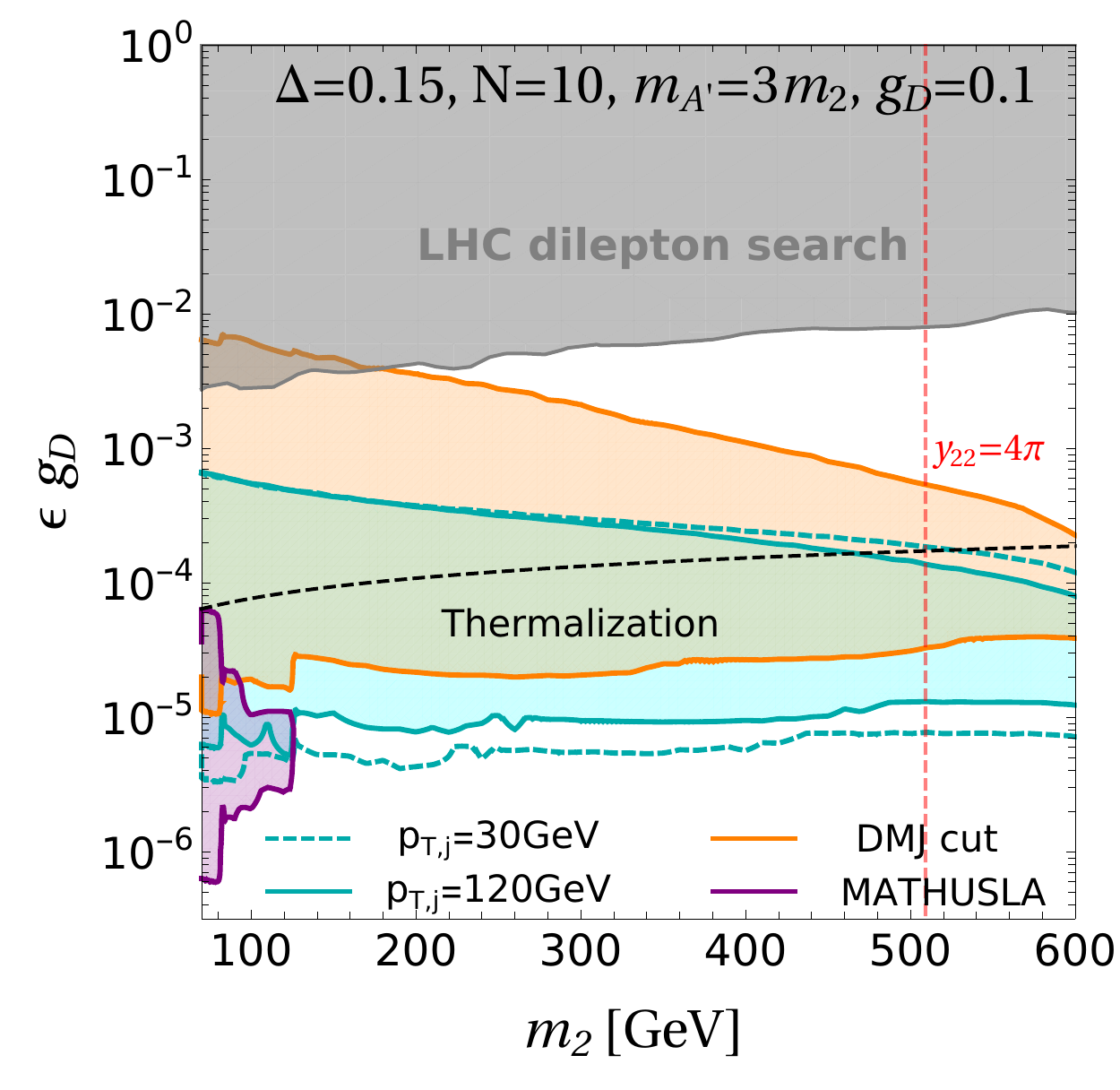}
	\caption{The expected sensitivity at HL-LHC to the scalar-vector model in the $\epsilon g_D$, $m_2$ plane for $\mathcal L=3~{\rm ab}^{-1}$ and $\sqrt{s}=13$ TeV. We have three panels from left to right for $\Delta=0.05, \ \Delta=0.10$ and $\Delta=0.15$ respectively, and top (bottom) panels for the signal event number $N_{\rm sig}^{\ell \ell} = 2.3$ (10) respectively. The heavy dark photon mass is set as $m_{A'}=3m_2$ and we assume $g_D=0.1$. The constraints from LHC dilepton searches are plotted in gray shaded region. For LLPs search, the projected reach for time delay strategy with ISR jet $p_T^j>120\rm{GeV} (30\rm GeV)$ and  are shown as cyan solid (dashed) contours respectively. The orange (purple) contours show the projected reach for DMJ strategy (MATHUSLA detector)respectively. The black dashed contours indicate the thermal equilibrium condition. The red dashed contours show the constraints from the perturbation condition. }
	\label{fig:VecRes}
\end{figure}

The signal event number for $s_2$ decay  that satisfying the selection criteria can be expressed as
\begin{equation}
 N_{\rm sig}^{\ell \ell}=\mathcal L\cdot \sigma_{\rm sig}\cdot P(s_2) \cdot \epsilon_{cut},
  \label{eq:Nsigll}
\end{equation}
where $P(s_2)$ is the $s_2$ decay possibility inside the decay volume, $\mathcal{L}=3 {\rm ab^{-1}}$ is the integrated luminosity and $\epsilon_{cut}$ is the total cut efficiency.
We use the Monte Carlo simulation to determine the decay time of $s_2$ according to its momentum direction and lifetime, then fix the location of decay vertex ($r_{s_2}$ and $z_{s_2}$) and finally calculate the parameters $\Delta t_\ell$ according to the kinematics of $s_2$ and $\ell$.

Based on the three cut conditions listed above, we show the sensitivities for three search strategies in Fig.~\ref{fig:VecRes} for $\Delta = 0.05, \ 0.10 $ and $0.15$, with signal events reaching $N=2.3$ and $10$.
We can see that the timing search strategy has better reach for smaller $\epsilon$ than DMJ strategy. Because it prefers longer life-time comparing to DMJ method. For the optimistic leading jet $p_T$ cut (dashed cyan), the sensitivities increase significantly comparing with the conservative $p_T$ cut. For the DMJ method, it is subject to the requirement that $s_2$ decays inside the tracker system, which prefers larger $\epsilon$.
At the same time, as stated before, $\sigma(pp\to jA')$ is proportional to $\epsilon^2$, so when fixing $g_D=0.1$ larger $\epsilon g_D$ will induce larger cross section of $A'$ resonance.
The sensitivity at LHC will cover the region from $\epsilon g_D= 10^{-2}$ to $10^{-4}$ combing these two strategies for $m_2$ around 100--500 GeV. For heavier mass, the $A'$ is too heavy to produce on-shell, thus the sensitivities are greatly suppressed. For MUTHUSLA search, it is not as sensitive as the two methods at ATLAS and CMS. It is because the MATHUSLA detector requires longer decay length $\sim 100$ m and a smaller angular volume. Therefore, $s_2$ can arrive at the decay volume with a lower possibility, especially for heavy $s_2$.

In Fig.~\ref{fig:VecRes}, there is a dip at $m_2=m_Z$, because the sudden drop of $\lambda_{22}$ at $m_2=m_Z$, which is also the reason for the island in the MATHUSLA search.
Moreover, we compare the sensitivities between $\Delta=0.05$, $\Delta=0.10$ and $\Delta=0.15$. The sensitivity for $\Delta=0.15$ is generally better than the case of $\Delta=0.05$ and $\Delta=0.10$ when mass is same, because larger $\Delta$ will require larger $\lambda_{22}$ for the DM relic abundance.
Thus, it results in a larger cross section for $pp\to j s_2 s_2$. Furthermore, larger $\Delta$ will lead to more energetic decay products from $s_2$, which helps signal to pass the cut conditions.

It is worth mentioning that when $m_1 \Delta\gtrsim m_Z$, there will be $s_2 \to s_1 Z$ decay open with
an on-shell $Z$. Although both suppressed by the factor $\epsilon^2 g_D^2$, it will be more significant in branching ratio comparing with $s_2 \to s_1 \ell^+ \ell^-$, because it is 2-body final state phase-space. It leads to an additional information that the invariant mass of the displaced lepton pair should be around mass of $Z$, which can help to further suppress the SM background and lower the requirement in the trigger\cite{Bae:2020dwf}. In our current strategies, the sensitivity region can not reach the region with $m_1 \Delta\gtrsim m_Z$. But with less stringent cuts and triggers, it may reach this region, then this invariant mass information can play a role.

\section{Conclusions}
\label{sec:conclusions}

The coannihilation mechanism of DM can be used to evade the direct detection constraints.
Usually, the coannihilation partner needs a sizable coupling to SM particles to obtain a large thermal cross section for the relic abundance. On the other hand, the coannihilation partner can be potentially long-lived at the detector scale, with the small coupling and mass splitting to the DM particle.
Previous studies mainly focus on the coannihilation between DM and the coannihilation partner,
which limits their mass to be lighter than 100 GeV.
In this work, we turn to the case that the coannihilation happens between the partner pair dominantly. This scenario opens heavy mass regions for DM and its coannihilation partner, and we focus on the collider searches for the long-lived coannihilation partner.

We introduced a generic model in which the DM candidate and its coannihilation partner are scalar particles, embedded in the iDM model. With the help of a broken symmetry, only the coannihilation partner couples to SM particles through a special Higgs portal coupling. The coannihilation partner pair annihilation dominates the DM effective annihilation cross section, while the DM-DM and DM-partner annihilation cross sections are negligible.  Next, we introduced
two specific models to illustrate how coannihilation partner can decay back to the DM particle and be long-lived.
The current limits from collider, indirect and direct searches are studied for the scenario and we propose to explore the model via the long-lived coannihilation partner.
We considered three methods here, namely displaced muon-jet method, timing method and MATHUSLA searches.
The first two methods utilized existing LHC detectors, ATLAS and CMS, together with appropriate triggers for the LLPs.
The basic cuts of triggers have been significantly relaxed by the presence of leptons in the partner decay final states.
The two methods shows good sensitivities for coannihilation partner with mass smaller than 500 GeV and kinetic mixing parameter $\epsilon$ between $10^{-1}-10^{-4}$ for $g_D = 0.1$. While the MATHUSLA search is less sensitive due to the small lifetime of the partner and the small angular decay volume.
In general, the LLP searches can provide a good sensitivity for the coannihilation DM scenario, which is complementary to the generic DM searches and can help to solve the mystery of the DM problem.

\section{Acknowledgments}
The work of JL is supported by National Science Foundation of China under Grant No. 12075005
and by Peking University under startup Grant No. 7101502597.
The work of XPW is supported by National Science Foundation of China under Grant No. 12005009.

\bibliographystyle{utphys}
\bibliography{ref}
\end{document}